\pdfoutput=1

\documentclass[%
twocolumn,
superscriptaddress,
 amsmath,amssymb,
 aps,
prd,
]{revtex4-1}

\usepackage{graphicx}
\usepackage{dcolumn}
\usepackage{bm}
\usepackage[T1]{fontenc}
\usepackage{appendix}
\usepackage[euler]{textgreek}
\usepackage{multirow}
\usepackage{xcolor}
\usepackage{hyperref}
\definecolor{darkblue}{rgb}{0.0,0.0,0.3}
\hypersetup{colorlinks,breaklinks,
            linkcolor=darkblue,urlcolor=darkblue,
            anchorcolor=darkblue,citecolor=darkblue}
\usepackage[mathlines]{lineno}
\setpagewiselinenumbers
\modulolinenumbers[5]

\begin{document}
\title{Improving sensitivity to low-mass dark matter in LUX using a novel electrode background mitigation technique}
\author{D.S.~Akerib} \affiliation{SLAC National Accelerator Laboratory, 2575 Sand Hill Road, Menlo Park, CA 94205, USA} \affiliation{Kavli Institute for Particle Astrophysics and Cosmology, Stanford University, 452 Lomita Mall, Stanford, CA 94309, USA} 
\author{S.~Alsum} \affiliation{University of Wisconsin-Madison, Department of Physics, 1150 University Ave., Madison, WI 53706, USA}  
\author{H.M.~Ara\'{u}jo} \affiliation{Imperial College London, High Energy Physics, Blackett Laboratory, London SW7 2BZ, United Kingdom}  
\author{X.~Bai} \affiliation{South Dakota School of Mines and Technology, 501 East St Joseph St., Rapid City, SD 57701, USA}  
\author{J.~Balajthy} \affiliation{University of California Davis, Department of Physics, One Shields Ave., Davis, CA 95616, USA}  
\author{J.~Bang} \affiliation{Brown University, Department of Physics, 182 Hope St., Providence, RI 02912, USA}  
\author{A.~Baxter} \affiliation{University of Liverpool, Department of Physics, Liverpool L69 7ZE, UK}  
\author{E.P.~Bernard} \affiliation{University of California Berkeley, Department of Physics, Berkeley, CA 94720, USA}  
\author{A.~Bernstein} \affiliation{Lawrence Livermore National Laboratory, 7000 East Ave., Livermore, CA 94551, USA}  
\author{T.P.~Biesiadzinski} \affiliation{SLAC National Accelerator Laboratory, 2575 Sand Hill Road, Menlo Park, CA 94205, USA} \affiliation{Kavli Institute for Particle Astrophysics and Cosmology, Stanford University, 452 Lomita Mall, Stanford, CA 94309, USA} 
\author{E.M.~Boulton} \affiliation{University of California Berkeley, Department of Physics, Berkeley, CA 94720, USA} \affiliation{Lawrence Berkeley National Laboratory, 1 Cyclotron Rd., Berkeley, CA 94720, USA} \affiliation{Yale University, Department of Physics, 217 Prospect St., New Haven, CT 06511, USA}
\author{B.~Boxer} \affiliation{University of Liverpool, Department of Physics, Liverpool L69 7ZE, UK}  
\author{P.~Br\'as} \affiliation{LIP-Coimbra, Department of Physics, University of Coimbra, Rua Larga, 3004-516 Coimbra, Portugal}  
\author{S.~Burdin} \affiliation{University of Liverpool, Department of Physics, Liverpool L69 7ZE, UK}  
\author{D.~Byram} \affiliation{University of South Dakota, Department of Physics, 414E Clark St., Vermillion, SD 57069, USA} \affiliation{South Dakota Science and Technology Authority, Sanford Underground Research Facility, Lead, SD 57754, USA} 
\author{M.C.~Carmona-Benitez} \affiliation{Pennsylvania State University, Department of Physics, 104 Davey Lab, University Park, PA  16802-6300, USA} 
\author{C.~Chan} \affiliation{Brown University, Department of Physics, 182 Hope St., Providence, RI 02912, USA}  
\author{J.E.~Cutter} \affiliation{University of California Davis, Department of Physics, One Shields Ave., Davis, CA 95616, USA}  
\author{L.~de\,Viveiros}  \affiliation{Pennsylvania State University, Department of Physics, 104 Davey Lab, University Park, PA  16802-6300, USA} 
\author{E.~Druszkiewicz} \affiliation{University of Rochester, Department of Physics and Astronomy, Rochester, NY 14627, USA}  
\author{A.~Fan} \affiliation{SLAC National Accelerator Laboratory, 2575 Sand Hill Road, Menlo Park, CA 94205, USA} \affiliation{Kavli Institute for Particle Astrophysics and Cosmology, Stanford University, 452 Lomita Mall, Stanford, CA 94309, USA} 
\author{S.~Fiorucci} \affiliation{Lawrence Berkeley National Laboratory, 1 Cyclotron Rd., Berkeley, CA 94720, USA} \affiliation{Brown University, Department of Physics, 182 Hope St., Providence, RI 02912, USA} 
\author{R.J.~Gaitskell} \affiliation{Brown University, Department of Physics, 182 Hope St., Providence, RI 02912, USA}  
\author{C.~Ghag} \affiliation{Department of Physics and Astronomy, University College London, Gower Street, London WC1E 6BT, United Kingdom}  
\author{M.G.D.~Gilchriese} \affiliation{Lawrence Berkeley National Laboratory, 1 Cyclotron Rd., Berkeley, CA 94720, USA}  
\author{C.~Gwilliam} \affiliation{University of Liverpool, Department of Physics, Liverpool L69 7ZE, UK}  
\author{C.R.~Hall} \affiliation{University of Maryland, Department of Physics, College Park, MD 20742, USA}  
\author{S.J.~Haselschwardt} \affiliation{University of California Santa Barbara, Department of Physics, Santa Barbara, CA 93106, USA}  
\author{S.A.~Hertel} \affiliation{University of Massachusetts, Amherst Center for Fundamental Interactions and Department of Physics, Amherst, MA 01003-9337 USA} \affiliation{Lawrence Berkeley National Laboratory, 1 Cyclotron Rd., Berkeley, CA 94720, USA} 
\author{D.P.~Hogan} \affiliation{University of California Berkeley, Department of Physics, Berkeley, CA 94720, USA}  
\author{M.~Horn} \affiliation{South Dakota Science and Technology Authority, Sanford Underground Research Facility, Lead, SD 57754, USA} \affiliation{University of California Berkeley, Department of Physics, Berkeley, CA 94720, USA} 
\author{D.Q.~Huang} \affiliation{Brown University, Department of Physics, 182 Hope St., Providence, RI 02912, USA}  
\author{C.M.~Ignarra} \affiliation{SLAC National Accelerator Laboratory, 2575 Sand Hill Road, Menlo Park, CA 94205, USA} \affiliation{Kavli Institute for Particle Astrophysics and Cosmology, Stanford University, 452 Lomita Mall, Stanford, CA 94309, USA} 
\author{R.G.~Jacobsen} \affiliation{University of California Berkeley, Department of Physics, Berkeley, CA 94720, USA}  
\author{O.~Jahangir} \affiliation{Department of Physics and Astronomy, University College London, Gower Street, London WC1E 6BT, United Kingdom}  
\author{W.~Ji} \affiliation{SLAC National Accelerator Laboratory, 2575 Sand Hill Road, Menlo Park, CA 94205, USA} \affiliation{Kavli Institute for Particle Astrophysics and Cosmology, Stanford University, 452 Lomita Mall, Stanford, CA 94309, USA} 
\author{K.~Kamdin} \affiliation{University of California Berkeley, Department of Physics, Berkeley, CA 94720, USA} \affiliation{Lawrence Berkeley National Laboratory, 1 Cyclotron Rd., Berkeley, CA 94720, USA} 
\author{K.~Kazkaz} \affiliation{Lawrence Livermore National Laboratory, 7000 East Ave., Livermore, CA 94551, USA}  
\author{D.~Khaitan} \affiliation{University of Rochester, Department of Physics and Astronomy, Rochester, NY 14627, USA}  
\author{E.V.~Korolkova} \affiliation{University of Sheffield, Department of Physics and Astronomy, Sheffield, S3 7RH, United Kingdom}  
\author{S.~Kravitz} \affiliation{Lawrence Berkeley National Laboratory, 1 Cyclotron Rd., Berkeley, CA 94720, USA}  
\author{V.A.~Kudryavtsev} \affiliation{University of Sheffield, Department of Physics and Astronomy, Sheffield, S3 7RH, United Kingdom}  
\author{E.~Leason} \affiliation{SUPA, School of Physics and Astronomy, University of Edinburgh, Edinburgh EH9 3FD, United Kingdom}  
\author{B.G.~Lenardo} \affiliation{University of California Davis, Department of Physics, One Shields Ave., Davis, CA 95616, USA} \affiliation{Lawrence Livermore National Laboratory, 7000 East Ave., Livermore, CA 94551, USA} 
\author{K.T.~Lesko} \affiliation{Lawrence Berkeley National Laboratory, 1 Cyclotron Rd., Berkeley, CA 94720, USA}  
\author{J.~Liao} \affiliation{Brown University, Department of Physics, 182 Hope St., Providence, RI 02912, USA}  
\author{J.~Lin} \affiliation{University of California Berkeley, Department of Physics, Berkeley, CA 94720, USA}  
\author{A.~Lindote} \affiliation{LIP-Coimbra, Department of Physics, University of Coimbra, Rua Larga, 3004-516 Coimbra, Portugal}  
\author{M.I.~Lopes} \affiliation{LIP-Coimbra, Department of Physics, University of Coimbra, Rua Larga, 3004-516 Coimbra, Portugal}  
\author{A.~Manalaysay} \affiliation{Lawrence Berkeley National Laboratory, 1 Cyclotron Rd., Berkeley, CA 94720, USA} \affiliation{University of California Davis, Department of Physics, One Shields Ave., Davis, CA 95616, USA} 
\author{R.L.~Mannino} \affiliation{Texas A \& M University, Department of Physics, College Station, TX 77843, USA} \affiliation{University of Wisconsin-Madison, Department of Physics, 1150 University Ave., Madison, WI 53706, USA} 
\author{N.~Marangou} \affiliation{Imperial College London, High Energy Physics, Blackett Laboratory, London SW7 2BZ, United Kingdom}  
\author{D.N.~McKinsey} \affiliation{University of California Berkeley, Department of Physics, Berkeley, CA 94720, USA} \affiliation{Lawrence Berkeley National Laboratory, 1 Cyclotron Rd., Berkeley, CA 94720, USA} 
\author{D.-M.~Mei} \affiliation{University of South Dakota, Department of Physics, 414E Clark St., Vermillion, SD 57069, USA}  
\author{J.A.~Morad} \affiliation{University of California Davis, Department of Physics, One Shields Ave., Davis, CA 95616, USA}  
\author{A.St.J.~Murphy} \affiliation{SUPA, School of Physics and Astronomy, University of Edinburgh, Edinburgh EH9 3FD, United Kingdom}  
\author{A.~Naylor} \affiliation{University of Sheffield, Department of Physics and Astronomy, Sheffield, S3 7RH, United Kingdom}  
\author{C.~Nehrkorn} \affiliation{University of California Santa Barbara, Department of Physics, Santa Barbara, CA 93106, USA}  
\author{H.N.~Nelson} \affiliation{University of California Santa Barbara, Department of Physics, Santa Barbara, CA 93106, USA}  
\author{F.~Neves} \affiliation{LIP-Coimbra, Department of Physics, University of Coimbra, Rua Larga, 3004-516 Coimbra, Portugal}  
\author{A.~Nilima} \affiliation{SUPA, School of Physics and Astronomy, University of Edinburgh, Edinburgh EH9 3FD, United Kingdom}  
\author{K.C.~Oliver-Mallory}\email[]{k.oliver-mallory@imperial.ac.uk} \affiliation{Imperial College London, High Energy Physics, Blackett Laboratory, London SW7 2BZ, United Kingdom} \affiliation{University of California Berkeley, Department of Physics, Berkeley, CA 94720, USA} \affiliation{Lawrence Berkeley National Laboratory, 1 Cyclotron Rd., Berkeley, CA 94720, USA}
\author{K.J.~Palladino} \affiliation{University of Wisconsin-Madison, Department of Physics, 1150 University Ave., Madison, WI 53706, USA}  
\author{C.~Rhyne} \affiliation{Brown University, Department of Physics, 182 Hope St., Providence, RI 02912, USA}  
\author{Q.~Riffard} \affiliation{University of California Berkeley, Department of Physics, Berkeley, CA 94720, USA} \affiliation{Lawrence Berkeley National Laboratory, 1 Cyclotron Rd., Berkeley, CA 94720, USA} 
\author{G.R.C.~Rischbieter} \affiliation{University at Albany, State University of New York, Department of Physics, 1400 Washington Ave., Albany, NY 12222, USA}  
\author{P.~Rossiter} \affiliation{University of Sheffield, Department of Physics and Astronomy, Sheffield, S3 7RH, United Kingdom}  
\author{S.~Shaw} \affiliation{University of California Santa Barbara, Department of Physics, Santa Barbara, CA 93106, USA} \affiliation{Department of Physics and Astronomy, University College London, Gower Street, London WC1E 6BT, United Kingdom} 
\author{T.A.~Shutt} \affiliation{SLAC National Accelerator Laboratory, 2575 Sand Hill Road, Menlo Park, CA 94205, USA} \affiliation{Kavli Institute for Particle Astrophysics and Cosmology, Stanford University, 452 Lomita Mall, Stanford, CA 94309, USA} 
\author{C.~Silva} \affiliation{LIP-Coimbra, Department of Physics, University of Coimbra, Rua Larga, 3004-516 Coimbra, Portugal}  
\author{M.~Solmaz} \affiliation{University of California Santa Barbara, Department of Physics, Santa Barbara, CA 93106, USA}  
\author{V.N.~Solovov} \affiliation{LIP-Coimbra, Department of Physics, University of Coimbra, Rua Larga, 3004-516 Coimbra, Portugal}  
\author{P.~Sorensen} \affiliation{Lawrence Berkeley National Laboratory, 1 Cyclotron Rd., Berkeley, CA 94720, USA}  
\author{T.J.~Sumner} \affiliation{Imperial College London, High Energy Physics, Blackett Laboratory, London SW7 2BZ, United Kingdom}  
\author{N.~Swanson} \affiliation{Brown University, Department of Physics, 182 Hope St., Providence, RI 02912, USA}  
\author{M.~Szydagis} \affiliation{University at Albany, State University of New York, Department of Physics, 1400 Washington Ave., Albany, NY 12222, USA}  
\author{D.J.~Taylor} \affiliation{South Dakota Science and Technology Authority, Sanford Underground Research Facility, Lead, SD 57754, USA}  
\author{R.~Taylor} \affiliation{Imperial College London, High Energy Physics, Blackett Laboratory, London SW7 2BZ, United Kingdom}  
\author{W.C.~Taylor} \affiliation{Brown University, Department of Physics, 182 Hope St., Providence, RI 02912, USA}  
\author{B.P.~Tennyson} \affiliation{Yale University, Department of Physics, 217 Prospect St., New Haven, CT 06511, USA}  
\author{P.A.~Terman} \affiliation{Texas A \& M University, Department of Physics, College Station, TX 77843, USA}  
\author{D.R.~Tiedt} \affiliation{University of Maryland, Department of Physics, College Park, MD 20742, USA}  
\author{W.H.~To} \affiliation{California State University Stanislaus, Department of Physics, 1 University Circle, Turlock, CA 95382, USA}  
\author{L.~Tvrznikova} \affiliation{University of California Berkeley, Department of Physics, Berkeley, CA 94720, USA} \affiliation{Lawrence Berkeley National Laboratory, 1 Cyclotron Rd., Berkeley, CA 94720, USA} \affiliation{Yale University, Department of Physics, 217 Prospect St., New Haven, CT 06511, USA}
\author{U.~Utku} \affiliation{Department of Physics and Astronomy, University College London, Gower Street, London WC1E 6BT, United Kingdom}  
\author{A.~Vacheret} \affiliation{Imperial College London, High Energy Physics, Blackett Laboratory, London SW7 2BZ, United Kingdom}  
\author{A.~Vaitkus} \affiliation{Brown University, Department of Physics, 182 Hope St., Providence, RI 02912, USA}  
\author{V.~Velan} \affiliation{University of California Berkeley, Department of Physics, Berkeley, CA 94720, USA}  
\author{R.C.~Webb} \affiliation{Texas A \& M University, Department of Physics, College Station, TX 77843, USA}  
\author{J.T.~White} \affiliation{Texas A \& M University, Department of Physics, College Station, TX 77843, USA}  
\author{T.J.~Whitis} \affiliation{SLAC National Accelerator Laboratory, 2575 Sand Hill Road, Menlo Park, CA 94205, USA} \affiliation{Kavli Institute for Particle Astrophysics and Cosmology, Stanford University, 452 Lomita Mall, Stanford, CA 94309, USA} 
\author{M.S.~Witherell} \affiliation{Lawrence Berkeley National Laboratory, 1 Cyclotron Rd., Berkeley, CA 94720, USA}  
\author{F.L.H.~Wolfs} \affiliation{University of Rochester, Department of Physics and Astronomy, Rochester, NY 14627, USA}  
\author{D.~Woodward} \affiliation{Pennsylvania State University, Department of Physics, 104 Davey Lab, University Park, PA  16802-6300, USA}  
\author{X.~Xiang} \affiliation{Brown University, Department of Physics, 182 Hope St., Providence, RI 02912, USA}  
\author{J.~Xu} \affiliation{Lawrence Livermore National Laboratory, 7000 East Ave., Livermore, CA 94551, USA}  
\author{C.~Zhang} \affiliation{University of South Dakota, Department of Physics, 414E Clark St., Vermillion, SD 57069, USA}
\collaboration{LUX Collaboration}
\date{\today}
\begin{abstract}
This paper presents a novel technique for mitigating electrode backgrounds that limit the sensitivity of searches for low-mass dark matter (DM) using xenon time projection chambers. In the LUX detector, signatures of low-mass DM interactions would be very low energy ($\sim$keV) scatters in the active target that ionize only a few xenon atoms and seldom produce detectable scintillation signals. In this regime, extra precaution is required to reject a complex set of low-energy electron backgrounds that have long been observed in this class of detector. Noticing backgrounds from the wire grid electrodes near the top and bottom of the active target are particularly pernicious, we develop a machine learning technique based on ionization pulse shape to identify and reject these events. We demonstrate the technique can improve Poisson limits on low-mass DM interactions by a factor of $2$--$7$ with improvement depending heavily on the size of ionization signals. We use the technique on events in an effective $5$~tonne$\cdot$day exposure from LUX's 2013 science operation to place strong limits on low-mass DM particles with masses in the range $m_{\chi}\in0.15$--$10$~GeV. This machine learning technique is expected to be useful for near-future experiments, such as LZ and XENONnT, which hope to perform low-mass DM searches with the stringent background control necessary to make a discovery.

\begin{description}
\item[PACS numbers] 95.35.+d, 14.80.Ly, 29.40.-n, 95.55.Vj 
\end{description}
\end{abstract}
\pacs{Valid PACS appear here}
\maketitle

\section{\label{sec:level1}Introduction}

Numerous astrophysical observations suggest $\sim$25\% of the energy density of the universe is composed of a non-luminous, gravitationally interacting material known as dark matter (DM) \cite{Aghanim:2018eyx,Harvey:2015hha}. A class of weakly interacting massive particles (WIMPs) with masses $m_{\chi}\in10$~GeV$-10$~TeV is consistent with observational evidence \cite{Feng:2010gw}, but without confirmation via direct detection \cite{Cui:2017nnn,Akerib:2016vxi,Selvi:2018hby}, the community has begun to more seriously consider well-motivated, lower-mass alternatives \cite{2018BRN}.

The Large Underground Xenon (LUX) experiment has reported world-leading WIMP-nucleon scattering limits using 95 livedays of data from 2013 (WS2013) \cite{Akerib:2013tjd,Akerib:2015rjg,Akerib:2016lao,LUXMigdal2018} and final limits with increased exposure \cite{Akerib:2016vxi,Akerib:2017kat}. Although LUX is most sensitive to $m_{\chi}\gtrsim5$~GeV, there are analysis techniques that can be used to reduce the threshold, allowing us to search for low-mass dark matter using the existing datasets \cite{Akerib:2019zrt,Akerib:2018hck,Angle:2011th}.

This paper presents a new analysis of WS2013 data, utilizing the smallest signals recorded by the instrument. This is done by incorporating events that contain only ionization signals, which remain robust at very low energies where there are usually no detectable scintillation signals. While this approach improves LUX's sensitivity to $m_{\chi}\in0.15$--$5$~GeV, it introduces a complex set of low-energy backgrounds that have been observed in similar analyses in this class of detectors \cite{Angle:2011th,XENON100S2o2016,XENON1TS2o2019,Agnes:2018ves}. Recently, much progress has been made to characterise these backgrounds (see e.g.~\cite{Bailey:2016pnn,Tomas:2018pny,Akerib:2020jud} and references therein), but it remains the case that advanced analysis techniques will be required to mitigate them adequately. In this paper, we utilize a unique pulse-shape-based machine learning technique to address the most pernicious background, namely electron pulses originating from the grids. We discuss how this might be further improved upon in near future experiments, such as LZ and XENONnT, to aid a possible low-mass dark matter discovery.

\bigskip 

\begin{figure}[ht]
  \includegraphics[width=0.7\linewidth]{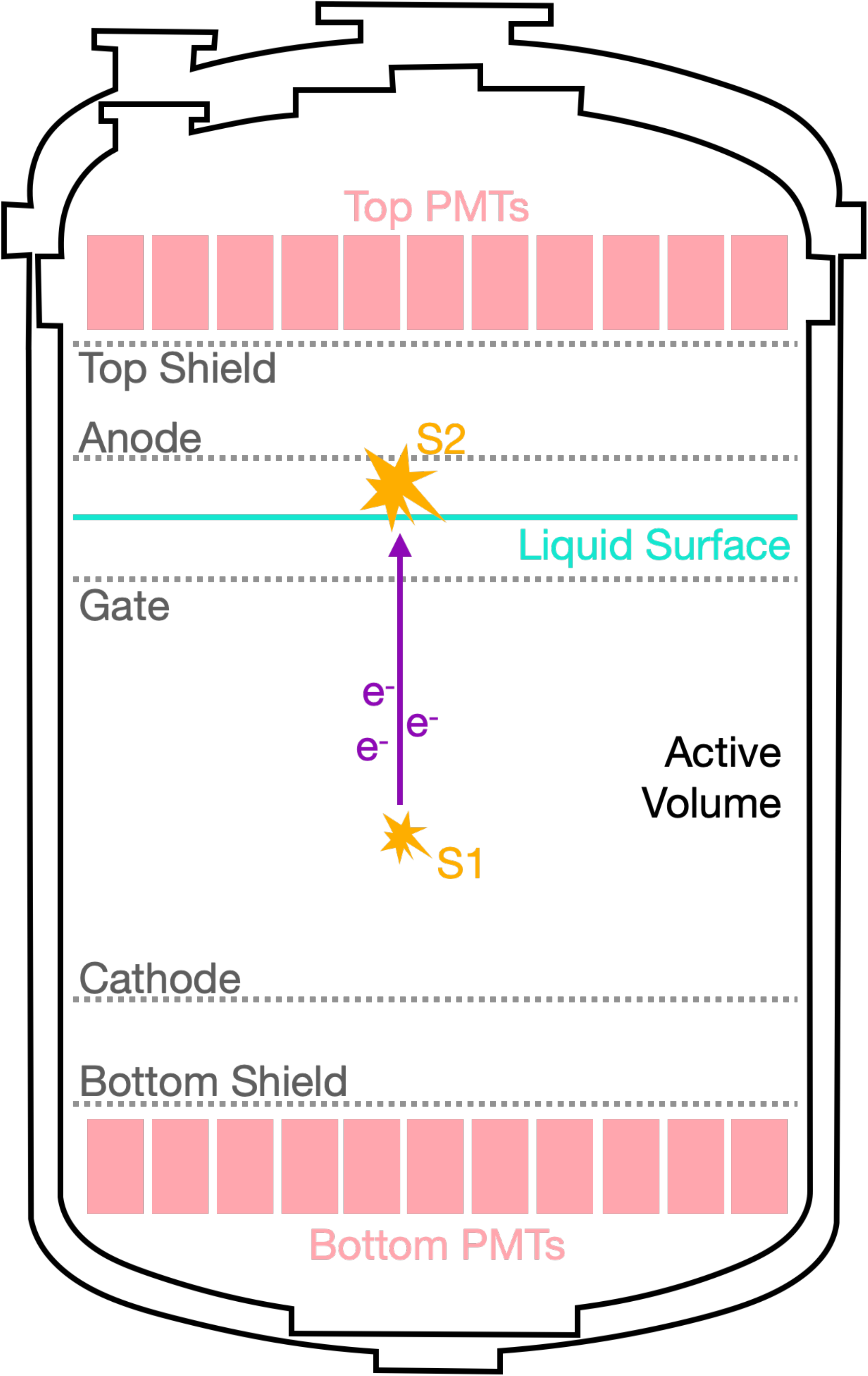}
  \caption{ A schematic illustration of the LUX detector.}
  \label{fig:det}
\end{figure}

The LUX detector, now retired, was a two-phase, xenon time projection chamber (TPC) that was operated at the Sanford Underground Research Facility (SURF). Particles scattering in the $250.9\pm2.1$~kg \cite{Akerib:2017vbi} of active liquid xenon mass can produce scintillation photons and free atomic electrons that are converted into two signals called $S1$ and $S2$. The $S1$ is created when the prompt scintillation is detected by 122 photomultiplier tubes (PMTs) located above and below the xenon, as illustrated in Fig.~\ref{fig:det}. The S2 is formed when the electrons drift upward in an electric field produced by gate and cathode wire grid electrodes near the top and bottom of the liquid. A stronger electric field across the liquid-vapor interface extracts the electrons into the gas phase creating proportional scintillation (electroluminescence) that is also detected by the PMTs \cite{Akerib:2012ys}. With an average amplification of $24.5$~photons detected (phd) for every electron, the detector was sensitive to single electrons \cite{Akerib:2020jud}.

LUX has powerful background rejection capability, because of its two signal readout. Radiogenic backgrounds occurring primarily near the edges of the xenon can be cut based on their three dimensional position. The horizontal coordinates of interaction vertices are reconstructed from the $S2$ hit pattern in the top PMTs ($x_{S2}$, $y_{S2}$), and the vertical coordinates are calculated from the product of drift velocity ($v_{drift}=0.152\pm0.001$~cm/$\mu$s \cite{Akerib:2017vbi}) and time delay between $S1$ and $S2$ ($t_d$). Additionally, discrimination between different types of incident particles is possible using the $S2$/$S1$ ratio. For example, $\beta$ particles and $\gamma$-rays scatter primarily on atomic electrons, producing relatively less scintillation and more electrons than DM scatters on nuclei for a given observed signal size \cite{Aprile:2006kx,Dahl:2009nta}. In the graphical space defined by $S2/S1$ and $S1$, the result is a distinguishable electron recoil (ER) band appearing above a nuclear recoil (NR) band.

In this work, we lower the LUX energy threshold close to the limit of the instrument by accepting events with only an $S2$, in addition to those with both an $S1$ and $S2$. The impact of this choice is illustrated in Fig.~\ref{fig:Efficiency}, which compares the low $S1+S2$ and the high $S2$ detection efficiencies in the energy range of xenon nuclear recoils produced by low-mass DM. The curves are calculated using the Noble Element Simulation Technique (NEST) version 2.0.1 \cite{NESTv2} to simulate the liquid xenon microphysics of signal production and detector physics of signal collection. Below $6$~keV$_{nr}$, the $S1+S2$ detection efficiency tapers off because low light collection efficiency of the PMTs ($\sim10$\% averaged over the active volume) prevents the small scintillation signals of nuclear recoils in this energy regime from producing $S1$s that pass LUXs two PMT coincidence criteria (described further in Sec.~\ref{sec:data-selection}). When the requirement that events contain an $S1$ is dropped, efficiency remains high down to $1$~keV$_{nr}$. At this point, the small number of electrons produced by nuclear recoil events are sometimes lost due to capture by impurities while drifting (a $0$--$30$\% effect referred to as ``electron lifetime'') or remain trapped at the liquid surface due to the $49$\% extraction efficiency observed in WS2013. (Further details on the additional efficiency curves presented in Fig.~\ref{fig:Efficiency} are discussed in Sec.~\ref{sec:data-selection}.)

An important side effect of incorporating events with only an $S2$ is that background rejection based on $S2/S1$ ratio and $z$-coordinate becomes impossible. As a consequence there are elevated rates of low-energy background events from multiple sources. As observed in \cite{Akerib:2020jud}, we saw delayed emission of electrons captured by electronegative impurities. These backgrounds were most noticeable within second-wide time windows following large events that produced many electrons. We also identified grid electron emission and grid radiogenic backgrounds (referred to here as ``electrode events'') as being particularly prevalent. The former is a process that produces few-electron $S2$s whose intensities correlate with the electric field magnitude near grid wires \cite{Bailey:2016pnn,Tomas:2018pny}. These events were observed to occur at hotspots on the grids that intermittently emit electrons, and are also expected to occur in less conspicuous patterns continuously throughout the run. The latter are backgrounds from $^{238}$U/$^{232}$Th contamination inherent to grid wires and plate-out of $^{222}$Rn daughters on wire surfaces. Primarily, plate-out occurs during construction when components are exposed to air with typical quantities of $^{222}$Rn. However, it also occurs during operation when charged $^{222}$Rn daughters in the xenon drift along electric field lines that terminate on wire surfaces. Small amounts of these isotopes are continuously absorbed in the xenon during normal operation conditions. Specifically in LUX, they were also introduced during a $150$~Bq $^{222}$Rn injection \cite{Akerib:SurfaceRun}. After an initial plate-out event, short-lived daughters quickly decay away leaving only $^{210}$Pb, with a $22$-year half-life, and its two daughters $^{210}$Bi and $^{210}$Po. The most harmful decay products are emitted at low energies. They include $^{210}$Pb and $^{210}$Bi $\beta$ decays, and recoils of $^{206}$Pb nuclei following $^{210}$Po $\alpha$ decays in which the $\alpha$ particle travels into the wire.

To address the excess of electrode events, a machine learning technique is used to identify and reject events based on $S2$ pulse shape, which is observed to differ between $S2$s originating on the electrodes and those originating in the bulk xenon. We show this technique can be used to significantly improve limits in an analysis including $S1$+$S2$ and $S2$-only events. We suggest how this technique might be refined for near future experiments, such as LZ and XENONnT.

\section{Data Selection and Analysis}\label{sec:data-selection}
To achieve maximum DM sensitivity in an ``$S1$-agnostic'' analysis, we introduced new cuts and vetoes that address background challenges not present in previous analyses requiring both an $S1$ and $S2$, \cite{Akerib:2015rjg,Akerib:2016lao,Akerib:2017uem,LUXMigdal2018}. The event selection criteria, exposure, and software threshold are described in further detail in the number points below.

\begin{enumerate}
\item Candidate DM events are single scatters in the active xenon volume. They are selected with a set of criteria similar to those described in \cite{Akerib:2013tjd,Akerib:2015rjg,Akerib:2016vxi}. Waveforms were required to contain exactly one $S2$ preceded by one or zero $S1$s; $S1$s were required to have a two PMT coincidence to distinguish them from a single PMT dark count; and $S2$s were required to have greater than 55 spikes in their pulse waveform.

The last criterion is an $S2$ threshold defined using a variable called ``spike count'', an alternative measure of pulse size sometimes used in place of area in sparse pulses. It corresponds to approximately 2.2 detected electrons as measured using the area of the pulse. It allows waveforms to contain additional small electron pulses from secondary photoionization and ionization phenomena that are sometimes induced by the primary event \cite{Akerib:2020jud}.

The single scatter detection efficiency, shown in Fig. \ref{fig:Efficiency}, was measured using tritium $\beta$ decay calibration data as a robust sample of low-energy events in the liquid xenon bulk \cite{Akerib:2015wdi}.

\begin{figure}[ht]
  \includegraphics[width=\linewidth]{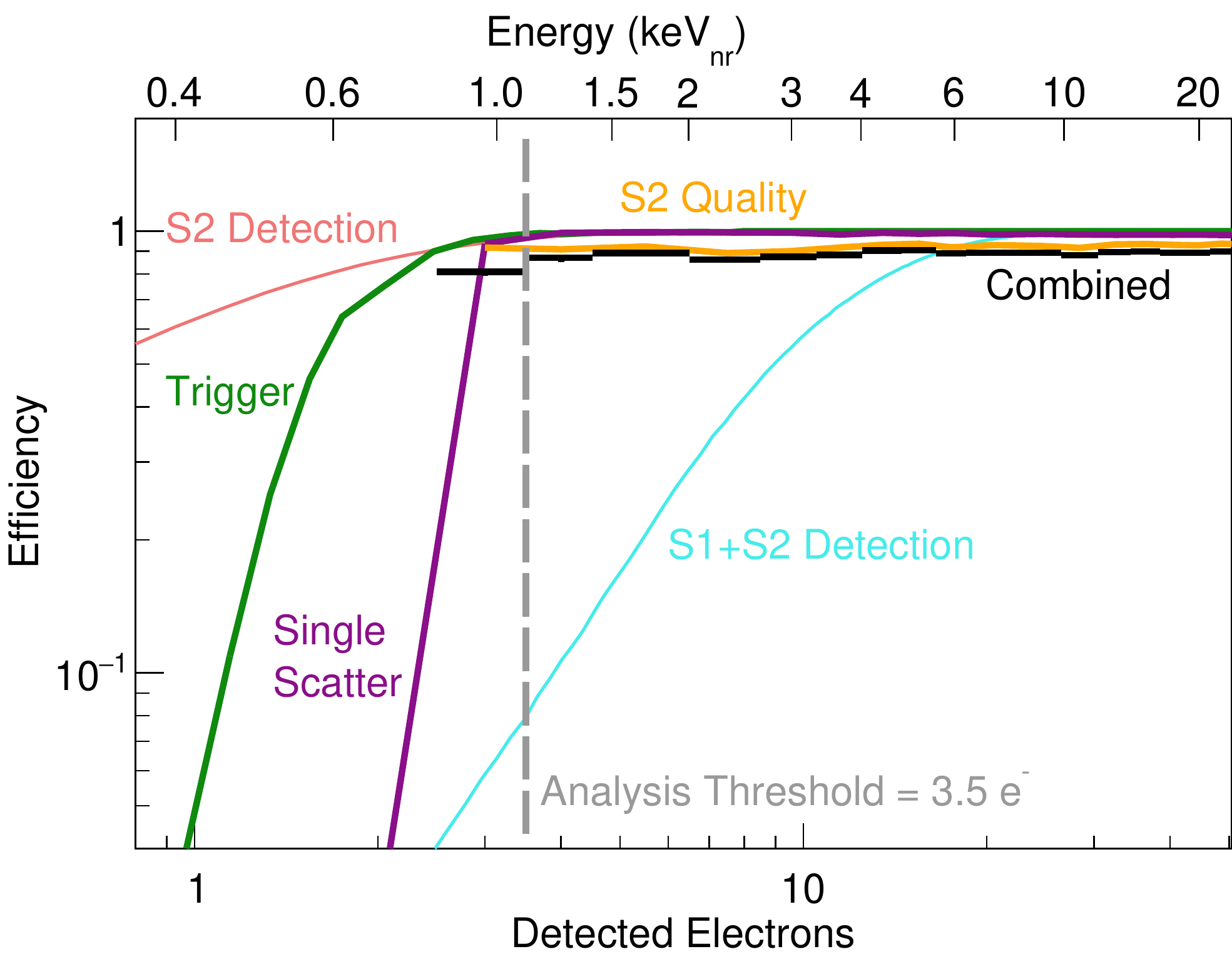}
  \caption{ Trigger \cite{Akerib:2018jcq}, single scatter, and $S2$-quality cut efficiencies, as well as their combined efficiency (including the $1.6$\% acceptance loss from the single photon $S1$ cut). The curves labeled ``$S2$ Detection'' and ``$S1$+$S2$ Detection'' encompass the liquid xenon microphysics of signal production and detector physics of signal collection as modeled with NEST v2.0.1 \cite{NESTv2}. The latter tapers off more quickly at low energy due to LUXs requirement that $S1$s be composed of photon signals in two or more PMT channels. It is not applied in this analysis, but is shown to illustrate the extra low-mass dark matter sensitivity gained in this search.}
  \label{fig:Efficiency}
\end{figure}

\item An $S2$ software threshold of $3.5$ detected electrons ($S2=85.75$~phd) was selected because efficiency measurements that use tritium calibration data are robust above this point. Below $3.5$ detected electrons, the calibration data was found to contain non-negligible quantities of background events from delayed emission of electrons by impurities.

\item Backgrounds from radioactive contamination on the electrodes and detector walls were mitigated with cuts on event position.
\begin{enumerate}
    \item Electrode events with $S1$s were removed with a drift time cut at $7<t_d<321~\mu$s, whose boundaries correspond to $5$~mm below the gate and $2$~mm above the cathode.
    \item Due to the geometry of the detector, $S2$s produced at the junction of the wall and cathode, a starting radius of $\sim24$~cm, drift upward with a slight radial bias exiting the liquid at a reconstructed radial coordinate of $r_{S2}\sim20$~cm \cite{Akerib:2017btb}. At this exit radius, $S2$s just above the software threshold have an uncertainty of $\sigma_{r_{S2}}\sim1.5$~cm \cite{Akerib:2017riv}. A cut of $r_{S2}<16$~cm, greater than $2\cdot\sigma_{r_{S2}}$ in from the exit radius, was selected to remove the vast majority of these events.
\end{enumerate}
The $124$~kg of xenon mass delimited by the two cuts was calculated from the fraction of accepted $^{83m}$Kr events (a calibration source that distributes itself uniformly in the active mass when injected into the detector). The uncertainty on the calculation has two major contributors. The first is a $\pm0.8$\% contribution from estimates of xenon volume and density used in the active mass calculation from \cite{Akerib:2017vbi}. The second is introduced by the drift cut which can only be applied to events that do have an $S1$ and, therefore, a determinate $t_d$. We estimate an additional $2$\% contribution to the uncertainty from the mass increase that would occur if the drift cut were to be removed.

\item Three data quality vetoes were used to remove WS2013 data taken under unstable detector conditions or during periods of time with high rates of $S2$-only backgrounds.
\begin{enumerate}
	\item Data taken during April 2013, the beginning of the WS2013 dataset, was removed from the analysis because of unstable electron lifetime during this period of time.
	\item A $50$~ms veto following large events with a full waveform area greater than $10^5$~phd was implemented to reduce backgrounds from delayed emission of electrons captured by impurities \cite{Akerib:2020jud}.
	\item Periods of time with an unusually high rate $S2$s of $2$--$3$ detected electrons (passing the radial cut and large event veto) were removed, because they were noticed to come from hotspots on the electrodes and correlate with high event rates above the threshold \cite{Bailey:2016pnn,Tomas:2018pny}.
\end{enumerate}
When combined, the three vetoes have the effect of reducing the livetime from the original $95$-day exposure to $81$~days.

\item The April 2013 dataset was scrutinized to identify several categories of pathological events. Multiple cuts on parameters derived by the LUX waveform processing software were designed to remove these events while maintaining $>90$\% efficiency as measured with tritium data.

\begin{enumerate}
    \item There were two types of pathological events associated with the extraction region between the gate and anode: those from interactions occurring in the vapor and those from interactions just below the liquid surface. These events produce proportional scintillation almost immediately after the energy deposition resulting in small or non-existent time delays between their $S1$s and $S2$s. They create $S2$-only backgrounds when the waveform processing software mistakenly identifies only a single $S2$ pulse from the combined $S1$+$S2$ trace. Because of the attached $S1$, these pulses often have bimodal shapes or reach a maximum very close to their beginning edge. They were tagged and removed using multiple cuts on a set of pulse shape defining parameters (these parameters are discussed in more detail in the context of electrode backgrounds in Sec. \ref{sec:electrode-backgrounds}).

    \item Many vapor events have wide, degraded $S2$s due to their electrons traveling through atypical, electroluminesence paths---not simply from liquid surface to anode. The waveform processing software tends to divided the waveforms of these events into many pulses labelling just one an $S2$. They were removed with a cut on ``badarea'', a parameter defined as the integrated area of the waveform trace less the area of the $S1$ and $S2$ pulses. To a smaller extent, the badarea cut removes backgrounds from delayed emission of electrons captured by impurities; however, the ``large event'' veto described in point 4.~(b) removes most of these backgrounds prior to application of this cut.
    \item Another fraction of the vapor events occur just above or below an anode grid wire causing an unusually large fraction of electroluminesence to reach the top or bottom PMTs. These events were mitigated with cut on the asymmetry of $S2$ light collected by top and bottom PMTs.
    \item A small fraction of events from radiocontamination on the detector walls have poor $x_{S2}$, $y_{S2}$ position reconstruction causing them to pass the radial cut defined in point 3.~(b). Because wall activities are $\sim10^2$ greater than the bulk, this pathology can be significant. These events were removed using a cut on the chi-squared value of the Mercury position reconstruction algorithm. Multiple scatters with interaction vertices at the same vertical position, but differing horizontal positions, are also removed via this cut.
\end{enumerate}

The effectiveness of the $S2$ quality cuts is quantified in Table \ref{tab:event_rate} which compares the combined acceptance of relevant vetoes/cuts with the event rate in the region just above the software threshold that most greatly determines sensitivity to low-mass DM.

\begin{table}[ht]
    \small
    \centering
    \caption{\label{tab:event_rate} Combined signal (tritium) acceptance of relevant vetoes/cuts compared with the percentage of events remaining just above the software threshold.}
    \begin{ruledtabular}
    \begin{tabular}{ r | c c}
    Description & Signal (Tritium) & Events \\
    of Cut/Veto & Acceptance & $3.5<n_e<4.5$ \\
    \colrule
    Starting Events & $100$\% & $100$\% \\
    Large Event Veto & $93$\% & $28$\% \\
    $S2$ Quality & $85$\% & $3.5$\% \\
    Electrode Hotspot Veto & $79$\% & $0.3$\% \\
    \end{tabular}
    \end{ruledtabular}
\end{table}

\item Events with $S2$/$S1$ ratios outside the $1$~\% and $99$~\% contours of the NR band were rejected for having the wrong recoil type.

\item Populations of $S2$-only events with a single photon pulse preceding the $S2$ by $0-7~\mu$s or $321-326~\mu$s are almost entirely composed of gate and cathode backgrounds that have $S1$s with too few photons to pass the two PMT coincidence requirement described in point 1. Because LUX's waveform processing software only records information about 10 pulses in each event some of the single photons are not recorded in the final dataset forcing us to identify and reject these backgrounds by eye. (Note that these single photon $S1$ pulses were the subject of a detailed study \cite{Akerib:2019zrt} and are well understood.)

Occasionally, the waveform of a genuine bulk event will contain a PMT dark count preceding the $S2$ by the aforementioned time windows, thus causing the event to be falsely identified as an electrode background. The efficiency loss due to this random coincidence was calculated to be $1.6$~\% using the dark count rate in \cite{Akerib:2019zrt}.
\end{enumerate}

\section{Machine Learning Technique for Removing Electrode Backgrounds}\label{sec:electrode-backgrounds}
\subsection{Mitigation with a Boosted Decision Tree}
\begin{figure}[ht]
  \includegraphics[width=\linewidth]{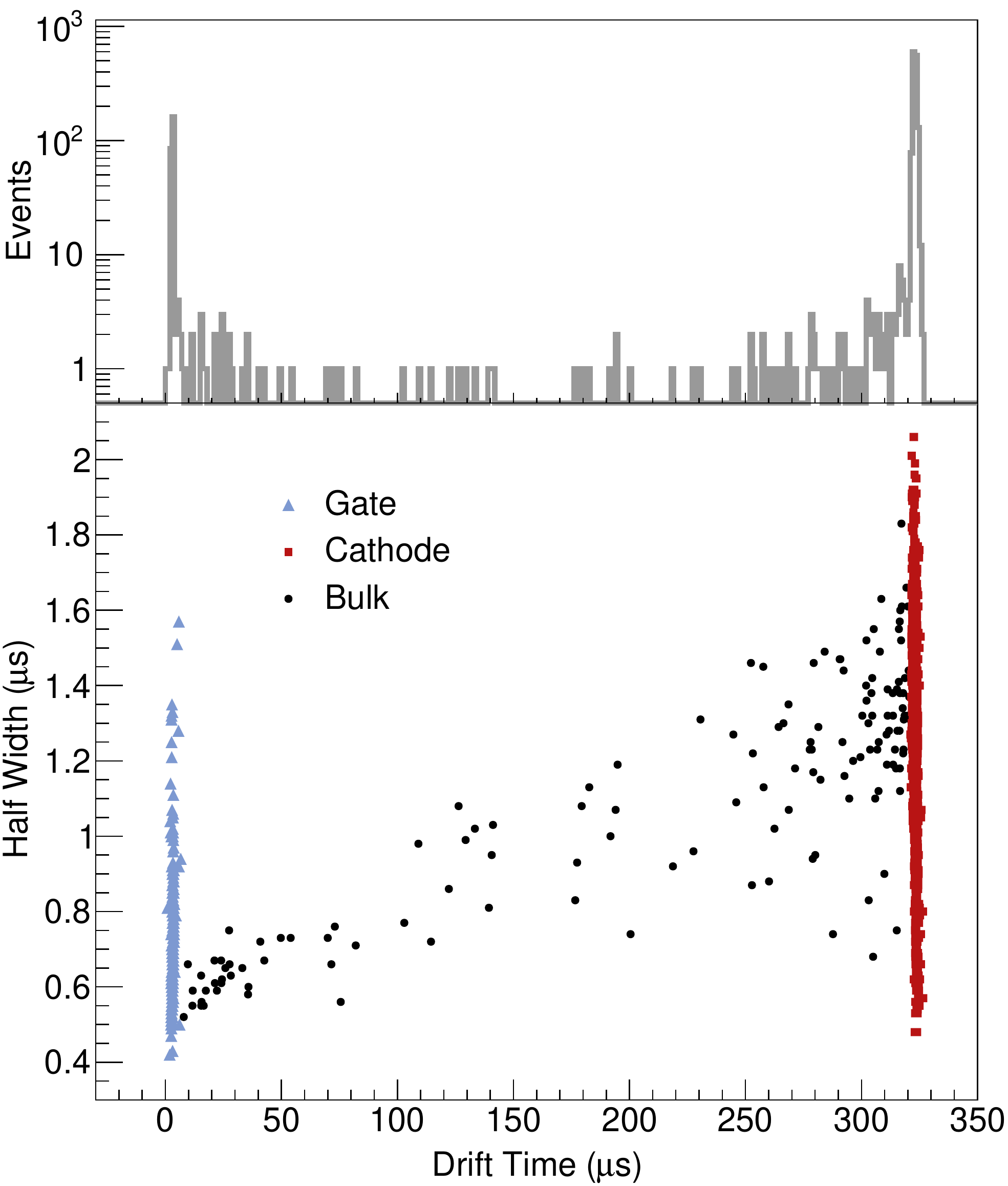}
  \caption{ All WS2013 events in range $3.5<n_e<50.5$ containing both an $S1$ (passing the two PMT coincidence requirement) and $S2$ pulse, and passing all vetoes/quality cuts applied to $S2$-only events. Gate, bulk, and cathode events are defined by drift time cuts: $t_d<7~\mu$s, $7~\mu$s~$<t_d<321~\mu$s, and $321~\mu$s~$<t_d$. The event rate is vastly higher at the gate and cathode drift times suggesting electrode events are the dominant source of backgrounds.}
  \label{fig:Width_Drift}
\end{figure}

The data selection criteria described in the previous section greatly reduce the background rate, but leave significant populations of events originating on the gate and cathode. This is evident in Fig.~\ref{fig:Width_Drift}~(top), which shows events in the dark matter search region that do have an $S1$ and, therefore, a determinate $z$ position.

Fig.~\ref{fig:Width_Drift}~(bottom) shows the $S2$ pulse width distributions of LUX's gate, bulk xenon, and cathode events defined by the drift time cuts $t_d<7~\mu$s, $7~\mu$s~$<t_d<321~\mu$s, and $321~\mu$s~$<t_d$. The bulk distribution shows a $\sqrt{t_d}$ trend that is consistent with expectations from diffusion of drifting electrons. If $S2$-only background events from interactions on the gate and cathode followed this trend, they could be removed using a pulse width cut. However, the gate and cathode distributions have a significantly broader spread than what is observed in the bulk, indicating that efficient removal of electrode backgrounds requires a more sophisticated cut. (Note that, in this analysis, we use a pulse half width parameter instead of full width. It is defined from the leading edge of the pulse, and avoids asymmetric tails occurring on the trailing edge of some pulses due to delayed emission of $S2$ electrons by mechanisms summarised in \cite{Akerib:2020jud}).

\begin{figure}[ht]
  \includegraphics[width=\linewidth]{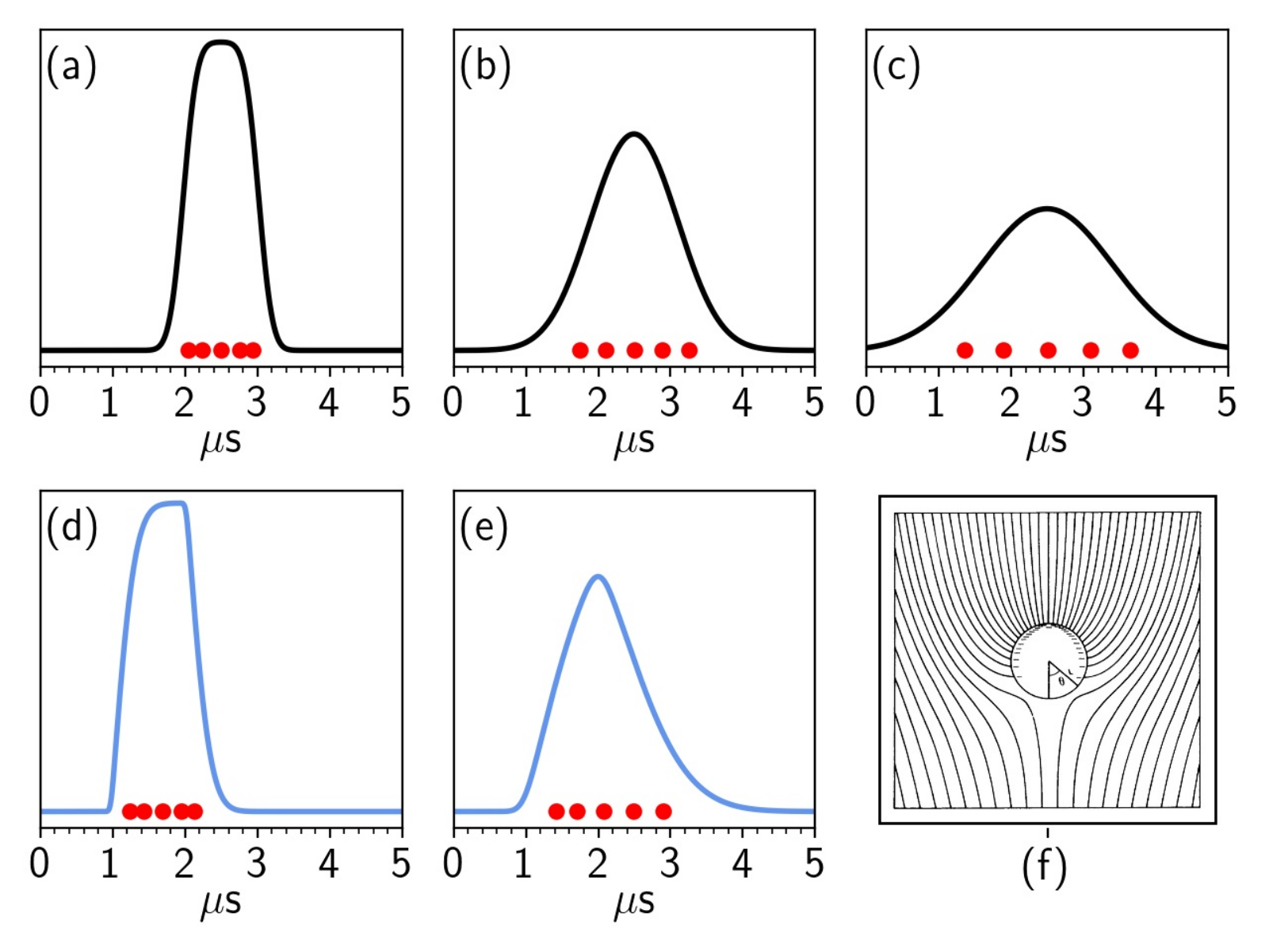}
  \caption{Panels (a--e) show typical $S2$ pulse shapes obtained from skew-Gaussian fits to LUX data. The top panels have symmetric shapes that are characteristic of bulk events near the top, middle, and bottom of the TPC (drift times $\sim10~\mu$s, $\sim150~\mu$s, and $\sim300~\mu$s). Panels (d--e) show asymmetry characteristic of gate $S2$s whose electrons travel through a range of electric fields. Panel (f) shows a typical electric field profile around a single gate wire (reproduced from \cite{Rolandi:2008qla}). Electric field profiles of cathode wires have a similarly wide range of magnitudes. The five red circles in panels (a-e) are parameters ($t_{10}$, $t_{25}$, $t_{50}$, $t_{75}$, and $t_{90}$) that mark the times at which the pulse attains $10$\%, $25$\%, $50$\%, $75$\%, and $90$\% of its total area. These parameters were used as input to the machine learning algorithm along with maximum pulse height, the time at which the pulse attains its maximum height, and the times at which the rising and falling edges of the pulse reach $0$~phd/sample. Note the bulk event profiles on the top panels can also occur for gate and cathode $S2$s originating on the top of a wire where field fringing is less dramatic. }
  \label{fig:waveform-examples}
\end{figure}

The discrepancy between electrode and bulk xenon $S2$ width distributions can be explained by the differences in pulse shape summarized in Fig.~\ref{fig:waveform-examples}. Panels (a--e) contain skew-Gaussian fits of $S2$ pulses typically found in the WS2013 data. The top panels are fits to bulk xenon $S2$s from the top, middle, and bottom of the TPC (drift times of $\sim10~\mu$s, $\sim150~\mu$s, and $\sim300~\mu$s). The $S2$s are symmetric and have widths that grow predictably with $\sqrt{t_d}$. Some electrode $S2$s are indistinguishable from those in the bulk xenon, but others are asymmetric and/or sharply peaked like the fits shown in panels (d--e). This variety of electrode $S2$ pulse shapes impairs the accuracy of typical pulse width metrics resulting in broader width distributions.

The phenomenon generating the odd $S2$ pulse shapes is electric field fringing around grid wires. Fig.~\ref{fig:waveform-examples}~(f) shows a typical electric field profile around a single gate wire (reproduced from \cite{Rolandi:2008qla}). The field lines do not point uniformly in the $z$ direction; instead, they stretch away from the wire surface in acurate patterns. Furthermore, very near the wire surface, the electric field magnitude is proportional to $\sim1/r_w$, where $r_w$ is the wire's radial coordinate. A 2-dimensional COMSOL model was used to calculate the electric field magnitude near gate and cathode wires in LUX. They are a factor of $\sim10^2$ greater than in the bulk xenon.

When an interaction occurs in a region with electric field fringing, it creates a cloud of electrons that is contorted such that it produces an odd $S2$ pulse shape. As the electrons drift they experience drastic changes in electric field that cause their velocity to vary and, thus, the distance between adjacent electrons to stretch or contract. Additionally, sufficiently far apart electrons can experience differences in path length that significantly alter their proximity. The latter effect is most significant near the bottom of gate wires where the curvature of field lines is most dramatic. Events near the cathode wires usually do not produce $S2$s, because their electrons drift downward along field lines that terminate on the electrode that shields the bottom PMTs. Only $16$\% of field lines near cathode wires extend upward and those that do stem from the top of the wire where field curvature is less dramatic. Despite this, $S2$s originating on the cathode, as well as those on the gate, can have visibly asymmetric shapes.

Fortunately, the odd $S2$ shapes created by field fringing can be successfully tagged by holistically quantifying pulse shape. For this reason, we used a machine learning algorithm to design a cut on the full set of LUX parameters that quantify shape. Some examples of these parameters are provided in Fig.~\ref{fig:waveform-examples}~(a--e). The five red circles along the horizontal axis are the points at which the pulse attains $10$, $25$, $50$, $75$ and $90$\% of its total area. The varying distances between red circles in panels (a--e) is evidence that pulse shape asymmetry is encoded in these quantities. Other shape quantifying parameters are the maximum pulse height, the time at which the pulse attains its maximum height, and the times at which the rising and falling edges of the pulse reach $0$~phd/sample.

Out of the many available machine learning methods, we choose a boosted decision tree (BDT) for ease in understanding how the algorithm uses and values input parameters. This algorithm is implemented in Python through the \textit{scikit-learn} library's AdaBoost (Adaptive Boosting \cite{scikit-learn}) classifier. The output is a discriminator score that indicates whether an $S2$ is more bulk- or electrode-like.

AdaBoost builds a sequence of weak classifiers that focus on different subsets of the training data. In our case, the weak classifiers are decision trees that perform cuts on input parameters to decide whether each datapoint should be assigned a value of $+1$ (bulk-like) or $-1$ (electrode-like). When the $n$th tree has been trained, the datapoints are re-weighted so that misclassified points are given higher importance during training of the ($n+1$)th tree. This iterative procedure is repeated until the desired number of trees have been trained. Although the prediction of each tree is binary, the final discrimination score may take on continuous values from $-1$ to $+1$. It is calculated by averaging the predictions of the decision trees with weights that vary depending on the classification accuracy of the tree \cite{adaboost}.

\begin{figure}[ht]
  \includegraphics[width=\linewidth]{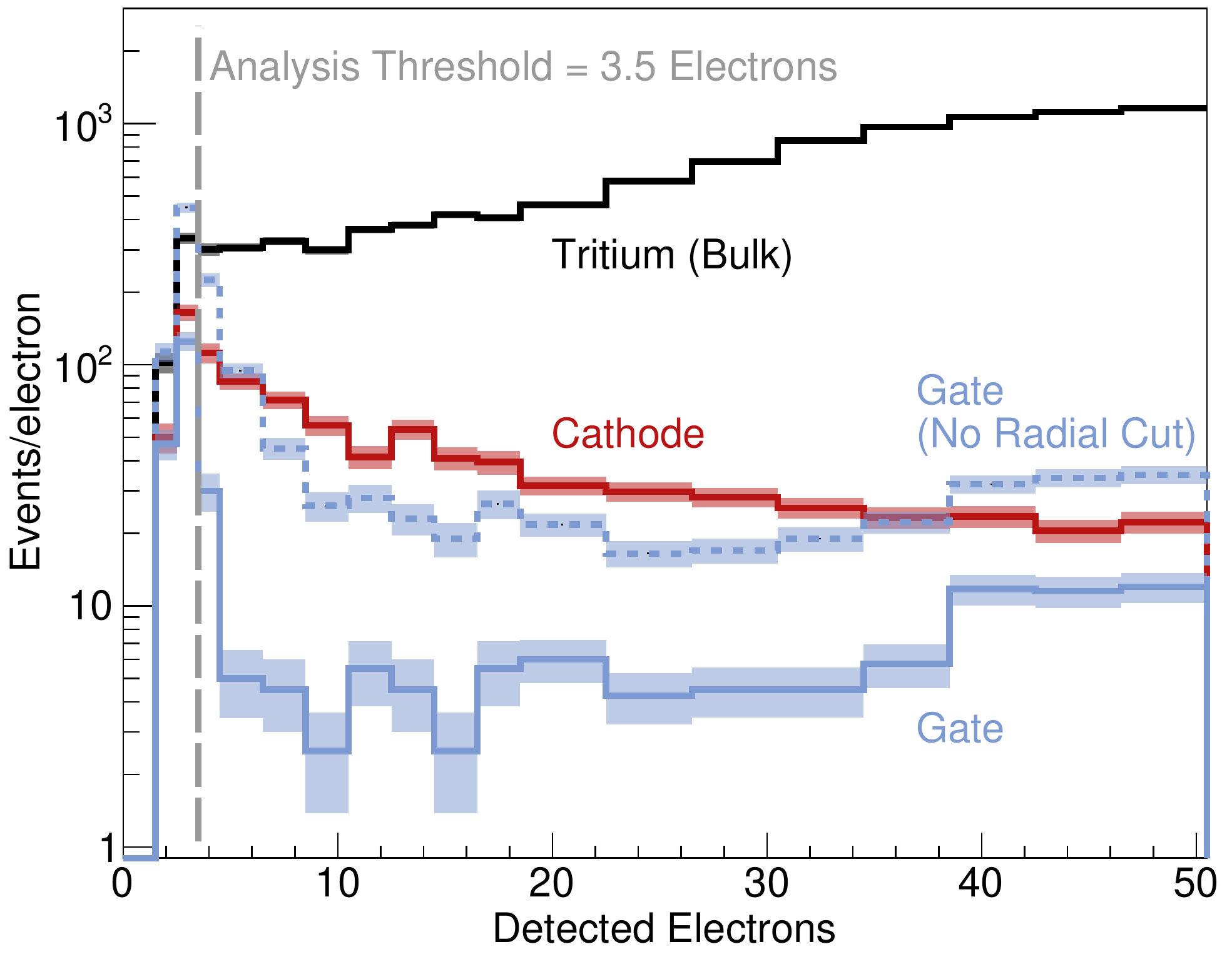}
  \caption{WS2013 gate, cathode, and tritium calibration data used to train a boosted decision tree to recognize electrode backgrounds. All events pass the $S2$-only quality cuts, except some gate events that fail the radial cut. This cut was removed to increase the number of gate training events. Before training, this data was re-weighted to give identical spectra for the three sources, as well as a 1:1 ratio of gate/cathode events.}
  \label{fig:Training_Data}
\end{figure}

The training and testing data were selected to include $S2$s of all sizes in the dark matter search range of $3.5-50.5$~detected electrons (Fig.~\ref{fig:Training_Data}). Bulk-like events were sourced from the tritium calibration dataset. They were required to pass all selection criteria outlined in Sec.~\ref{sec:data-selection}. Gate- and cathode-like events were sourced from WS2013. They were required to have both an $S1$ and $S2$, pass all vetoes/quality cuts applied to $S2$-only events, and pass gate or cathode drift time cuts of $t_d<7~\mu$s or $321~\mu$s~$<t_d$. Because these criteria produced many more cathode than gate events, the radial cut was relaxed for the latter, to generate a sufficiently large training sample. This choice could potentially make the training data less representative of the backgrounds in WS2013, but this only affects how optimally the BDT removes grid backgrounds and not how conservative the final limit is (which is determined by the Yellin procedure, Sec.~\ref{sec:results}).

Despite the electric field profile around cathode wires preventing most events from producing $S2$s, there are many more cathode than gate training events. This is not reliable evidence of greater $^{222}$Rn daughter contamination or a higher electron emission rate on the cathode. Instead, it is primarily a result of detector conditions that tend to enlarge $S2$s originating on the gate to sizes greater than the $50.5$ detected electron upper threshold used in this analysis. Primarily, the larger average electric field near gate wires ($52$~kV/cm compared to $18$~kV/cm on the cathode) enhances the charge yield. Additionally, electrons originating on the cathode drift through the full length of the detector over which they have a $30$\% chance of being captured by a xenon impurity, while those originating on the gate drift a very short distance and have a very low capture probability. This causes more cathode events to appear in the signal region below $50.5$ detected electrons.

The gate, cathode, and tritium training data were weighted to share the same flat $S2$ area spectrum with a single step at $12.5$ detected electrons. Below the step, where both the DM signal spectrum and the April 2013 background spectrum are strongest, the training data were weighted more heavily. Keeping the same spectrum for bulk- and electrode-like classes ensures the algorithm does not rely on $S2$ area as a means of discrimination, through residual correlation with input parameters. It is necessary because the shapes of the gate and cathode background spectra are unknown, due to the lack of $S1$s for most events.

Tests on the April 2013 data that were excluded from the final analysis, but pass the selection criteria outlined in Sec.~\ref{sec:data-selection}, were performed to find the ratios of gate:cathode:tritium events (scale factors multiplying the weighted $S2$ area spectra) that lead to near-optimal reduction in event rate at fixed signal efficiency. The selected gate:cathode and electrode (gate+cathode):tritium ratios were both 1:1.

Finally, cross-validation on multiple subsets of the training data was used to optimize the BDT's hyperparameters, such as maximum decision tree depth and number of trees.

The BDT training performance is best summarized with receiver operating characteristic (ROC) curves. These are plots of signal [tritium (bulk)] acceptance ($\epsilon_s$) vs electrode background leakage ($\epsilon_b$) calculated by placing a threshold cut on the discriminator scores of the testing datapoints at successive values spanning the range $(-1, +1)$. In this space, the diagonal $\epsilon_s=\epsilon_b$ corresponds to no discrimination (random guessing) and $\epsilon_s=1$ to perfect discrimination. The greater the area under the curve, the greater the discrimination power. 

\begin{figure}[ht]
  \includegraphics[width=\linewidth]{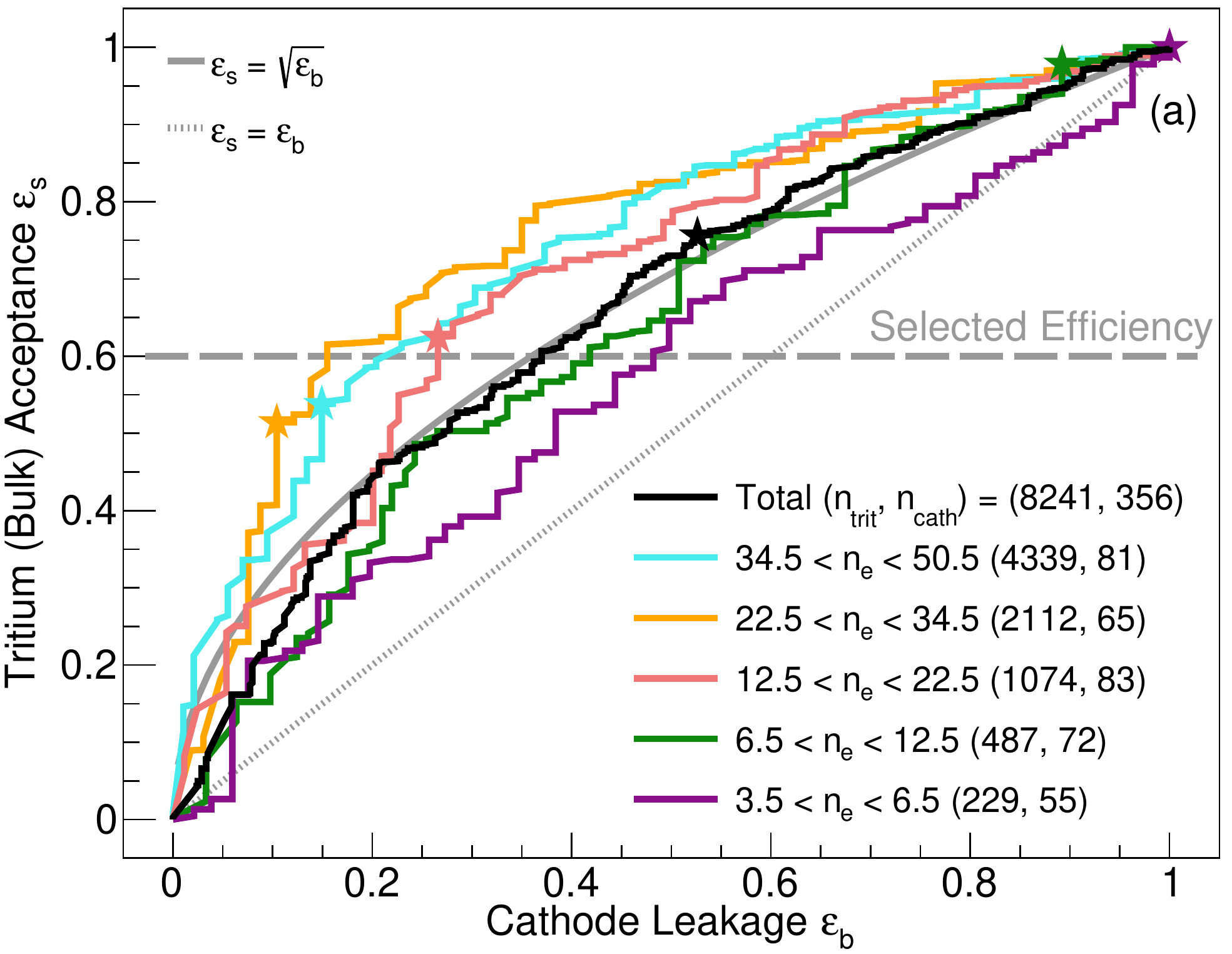}
  \includegraphics[width=\linewidth]{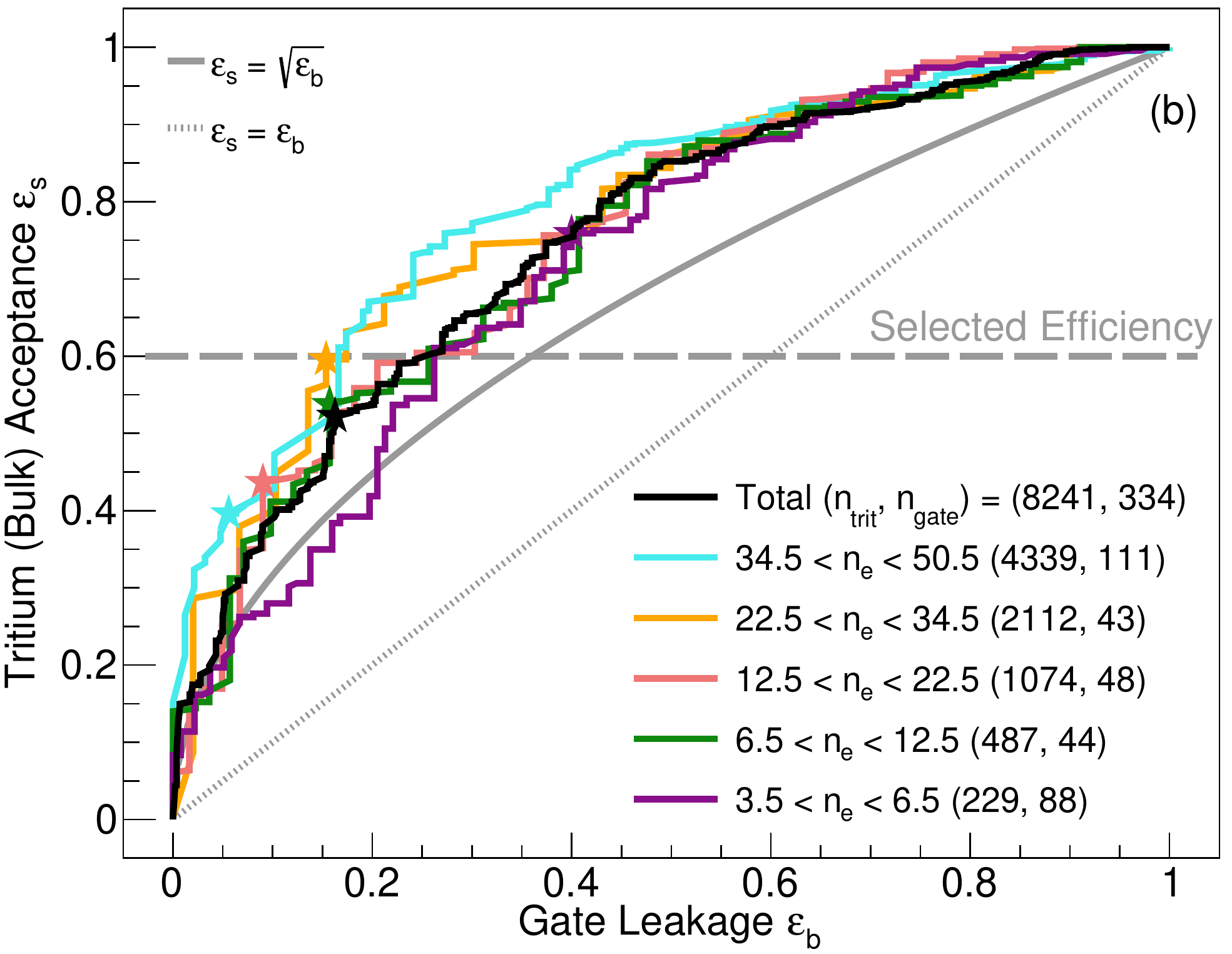}
  \caption{ROC curves for (a) bulk tritium vs cathode and (b) bulk tritium vs gate test data. The curves can be used to estimate a BDT discriminator threshold that maximizes sensitivity to dark matter signals in an extreme scenario where only grid backgrounds are present. Two additional curves are plotted to guide the eye: $\epsilon_s=\epsilon_b$ and $\epsilon_s=\sqrt{\epsilon_b}$, the thresholds that must be exceeded by the ROC curves for a BDT discriminator threshold to improve sensitivity in a Poisson or background subtracted DM analysis. Stars indicate the points of optimal $\epsilon_s/\sqrt{\epsilon_b}$.}
  \label{fig:roc}
\end{figure}

Families of ROC curves generated from gate and cathode training data with differing cuts on $S2$ size are shown in Fig.~\ref{fig:roc}. It is immediately clear the BDT is more adept at removing gate events compared to cathode events. This outcome was expected, as electron clouds from the cathode experience additional diffusion that erases initial $S2$ shape information. Discrimination power also increases with $S2$ size, because larger numbers of electrons form more recognizable pulse shape patterns. This trend is weaker for gate events, which can be efficiently identified and rejected even with few detected electrons.

The ROC curves can be used to estimate a BDT discriminator threshold that maximizes sensitivity to dark matter signals. To aid understanding of these curves, we consider two simple benchmark scenarios where the sensitivity of the analysis is a simple analytic function of the signal and background efficiencies, the quantities plotted in the curves. The background-subtracted scenario models a rare event search analysis with both an assumed signal and background model (e.g. a profile likelihood ratio). In this case, the sensitivity scales as $\epsilon_s/\sqrt{\epsilon_b}$, because the mean of the background spectrum can be subtracted, leaving only statistical fluctuations. The second scenario is a naive Poisson limit with no background subtraction, in which case the sensitivity scales as $\epsilon_s/\epsilon_b$. In order to improve the sensitivity relative to an analysis with no grid background cut, the ROC curves must exceed thresholds at $\epsilon_s=\sqrt{\epsilon_b}$ (background-subtracted) or $\epsilon_s=\epsilon_b$ (Poisson).

The present analysis uses Yellin's $p_\mathrm{max}$ test statistic \cite{Yellin:2002xd} to calculate sensitivity, as described further in Sec.~\ref{sec:results}. This approach is expected to scale somewhere between the Poisson and background subtracted extremes, because it takes into account the difference in shape of the data and signal energy distributions, but makes no assumptions about the shape of the background distribution. Given this, both $\epsilon_s=\sqrt{\epsilon_b}$ and $\epsilon_s=\epsilon_b$ thresholds are drawn in Fig.~\ref{fig:roc} for comparison with the gate and cathode ROC curves of each $S2$ size bin. All of the gate curves pass the more stringent $\epsilon_s=\sqrt{\epsilon_b}$ requirement for most values of the BDT discriminator threshold while the cathode curves pass the requirement only in bins with $S2$s of $n_e>12.5$. This behavior shows the BDT cut can be used to improve sensitivity in extreme scenarios where only gate backgrounds are present, or cathode backgrounds with $n_e>12.5$.

In our analysis, the discriminator cut was designed to be signal model agnostic by requiring signal acceptance be constant with respect to event energy. It was chosen considering the points of maximum $\epsilon_s/\sqrt{\epsilon_b}$, indicated as stars on the ROC curves. Based on the clustering of these points, a flat $60$\% signal acceptance was imposed by selecting a different discriminator threshold for each of the $S2$ size bins. At this signal acceptance, all ROC curves pass the $\epsilon_s/\sqrt{\epsilon_b}$ requirement except those from the cathode with $S2$s of $n_e<12.5$.

To evaluate the effectiveness of the discriminator cut, it was repeatedly modified to produce various choices of $\epsilon_s$ then applied to April 2013 data passing the selection criteria outlined in Sec.~\ref{sec:data-selection}. In each case, the observed data leakage matched the background leakage calculated from training data with 1:1 gate to cathode ratio, albeit subject to statistical fluctuations in a limited dataset. This result substantiates our conjecture that a significant fraction of the $S2$-only events remaining after the Sec.~\ref{sec:data-selection} cuts/vetoes are electrode backgrounds. Additionally, it is evidence that the gate and cathode training data is representative of the $S2$-only electrode backgrounds in that the $S2$s have similarly asymmetric/sharply peaked shapes.

\begin{figure}[ht]
  \includegraphics[width=\linewidth]{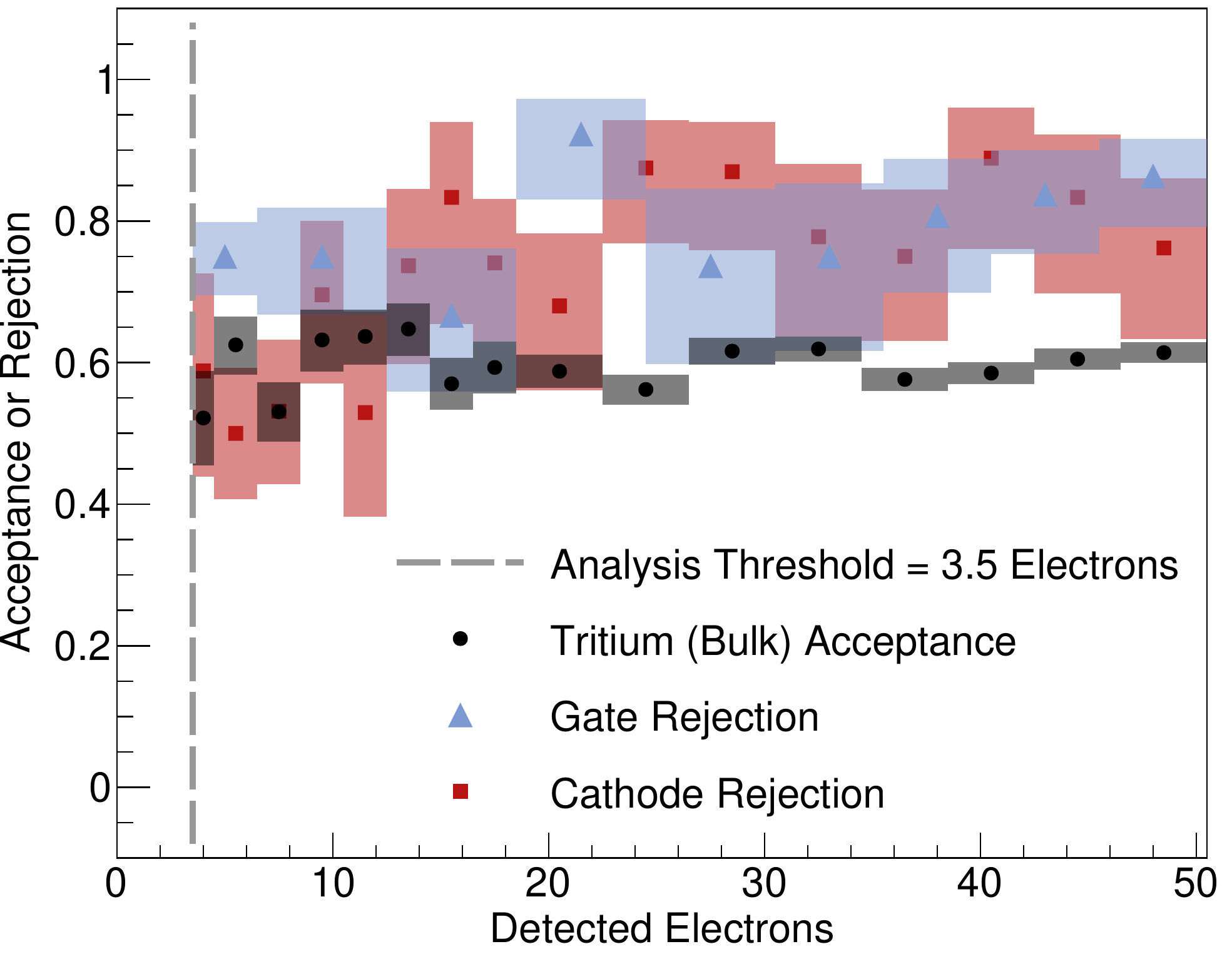}
  \caption{ Signal acceptance and background rejection capability of the boosted decision tree used to tag and remove gate- and cathode-like events. The cut is not applied below the software threshold.}
  \label{fig:Acceptance_Rejection}
\end{figure}

The signal acceptance and background rejection of the discriminator cut (calculated from the training data) are plotted in Fig.~\ref{fig:Acceptance_Rejection}. By design, the signal acceptance remains near $60$\% at all $S2$ sizes. The background rejection slightly improves with $S2$ size as is expected from the trends observed in the ROC curves. Additionally, gate and cathode rejection is similar for $S2$s with $n_e>12.5$, but cathode rejection falls short of gate rejection by $\sim15$\% for $n_e<12.5$.

\subsection{Effectiveness of Boosted Decision Tree}\label{sec:BDT-effectiveness}
An understanding of the parameters most useful to the BDT can be gained by looking at the ``importances'' of the input parameters presented in Tab.~\ref{tab:variable-ranking}. For a single decision tree, a parameter's importance is the fraction of training events separated by each branching node using the parameter, weighted by the reduction of impurity at each node. For a BDT, the overall importance of a parameter is an average of the individual decision tree importances with weights that vary depending on the classifcation accuracy of the tree (the same weights used in the calculation of the discriminator score). For our training data, the time differences outlining the middle of the pulse (e.g. $t_{75}-t_{50}$), as well as the pulse height, are more important than the time differences outlining the tails (not shown in Tab.~\ref{tab:variable-ranking}). This inclination of the BDT could be a result of $S2$ waveform noise from photoionization and ionization phenomena that more readily appears at the edges of $S2$ pulses. For example, the trailing edge of $S2$s can overlap with electron signals produced by the quantum efficiency of electroluminesence incident on the gate. These electrons will obscure the original shape of the $S2$'s trailing edge.

\begin{table}[ht]
\caption{\label{tab:variable-ranking}%
Importance of parameters in the BDT. Only the top five (of twelve) are tabulated, for brevity. Here, $t_x$ corresponds to the time at which the pulse attains $x$\% of its total area, as illustrated in Fig.~\ref{fig:waveform-examples}, while $V_{max}$ is the maximum pulse height. }
\begin{tabular}{p{2cm} r}
\hline \hline
\textrm{Variable} & \textrm{Importance}\\
\colrule
$t_{75}-t_{50}$ & 0.212\\
$V_{max}$ & 0.153\\
$t_{25}-t_{10}$ & 0.148\\
$t_{90}-t_{75}$ & 0.131\\
$t_{50}-t_{25}$ & 0.083\\
\hline \hline
\end{tabular}
\end{table}

\begin{figure}[ht]
  \includegraphics[width=\linewidth]{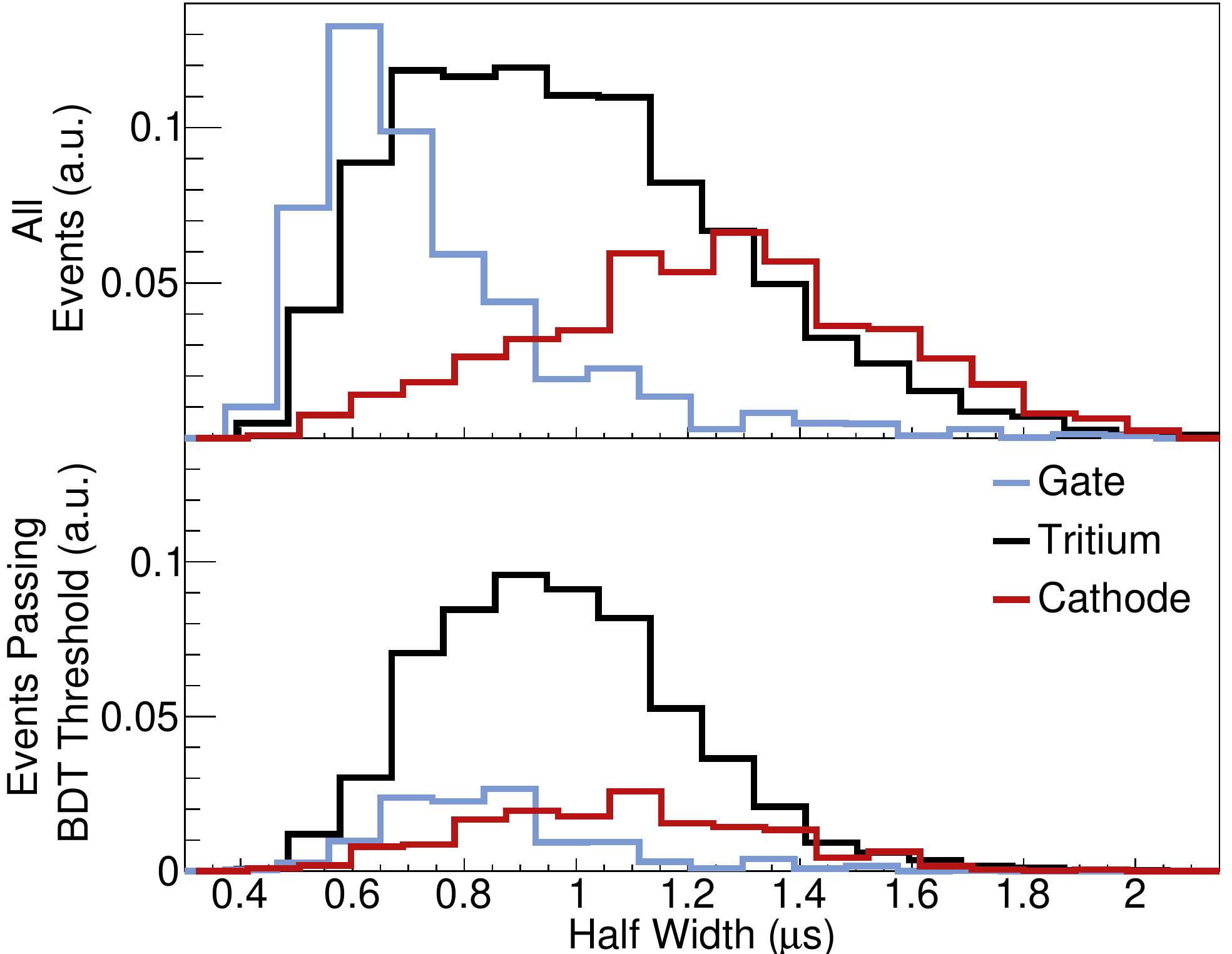}
  \caption{Half width distributions of training data before and after applying a discriminator cut tuned to $60$\% signal efficiency.}
  \label{fig:training_widths}
\end{figure}

Fig.~\ref{fig:training_widths} shows the half width distributions of training data before and after applying the selected discriminator cut. As designed, the cut greatly reduces the gate and cathode distributions while mostly preserving the tritium (bulk) distribution. While the averages of all three distributions shift towards central (bulk-like) values, there are tails remaining at small and large half widths. These tails are evidence the BDT is finding new features, more subtle than half width, that are useful for discriminating between electrode and bulk events.

\begin{figure}[ht]
  \includegraphics[width=\linewidth]{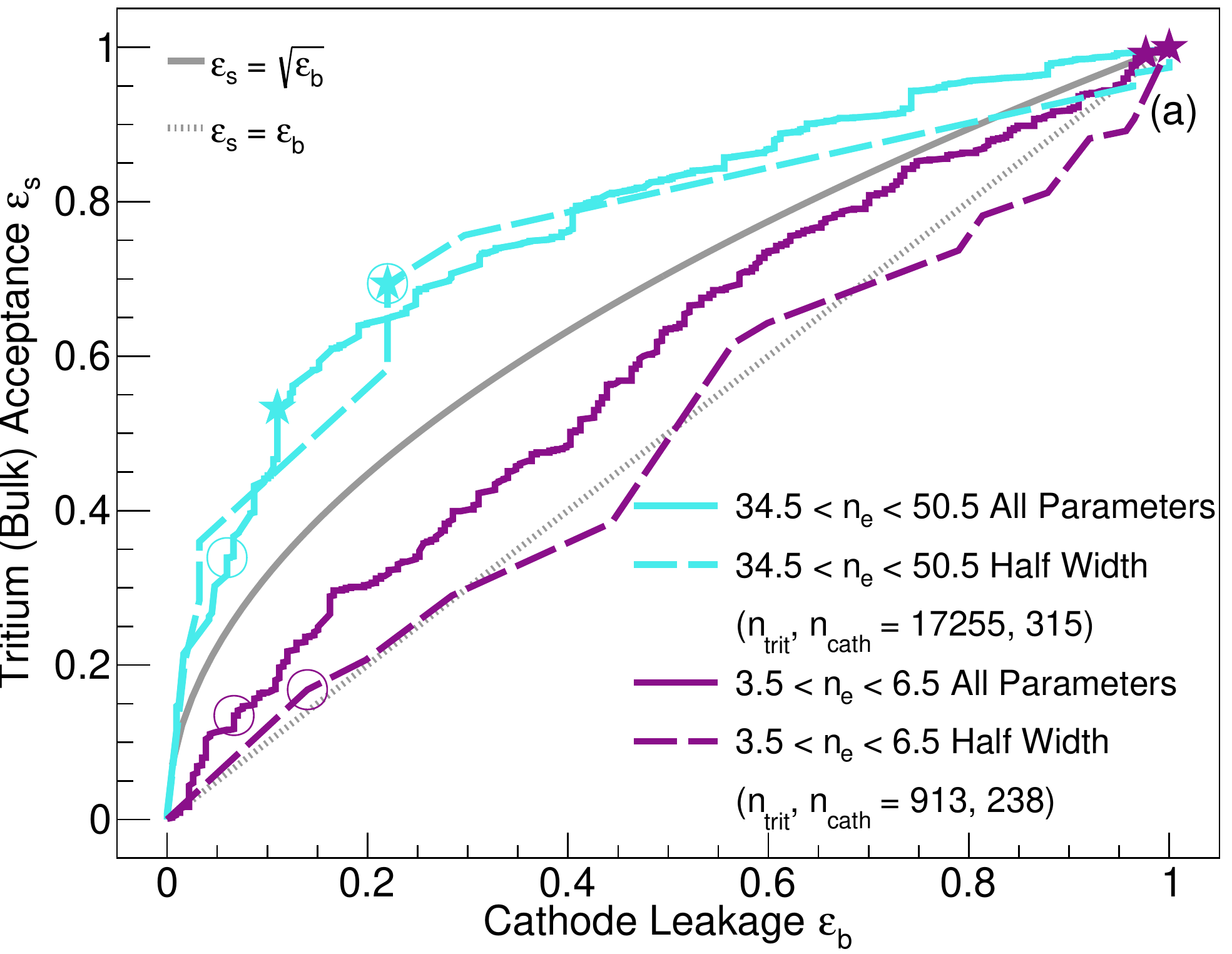}
  \includegraphics[width=\linewidth]{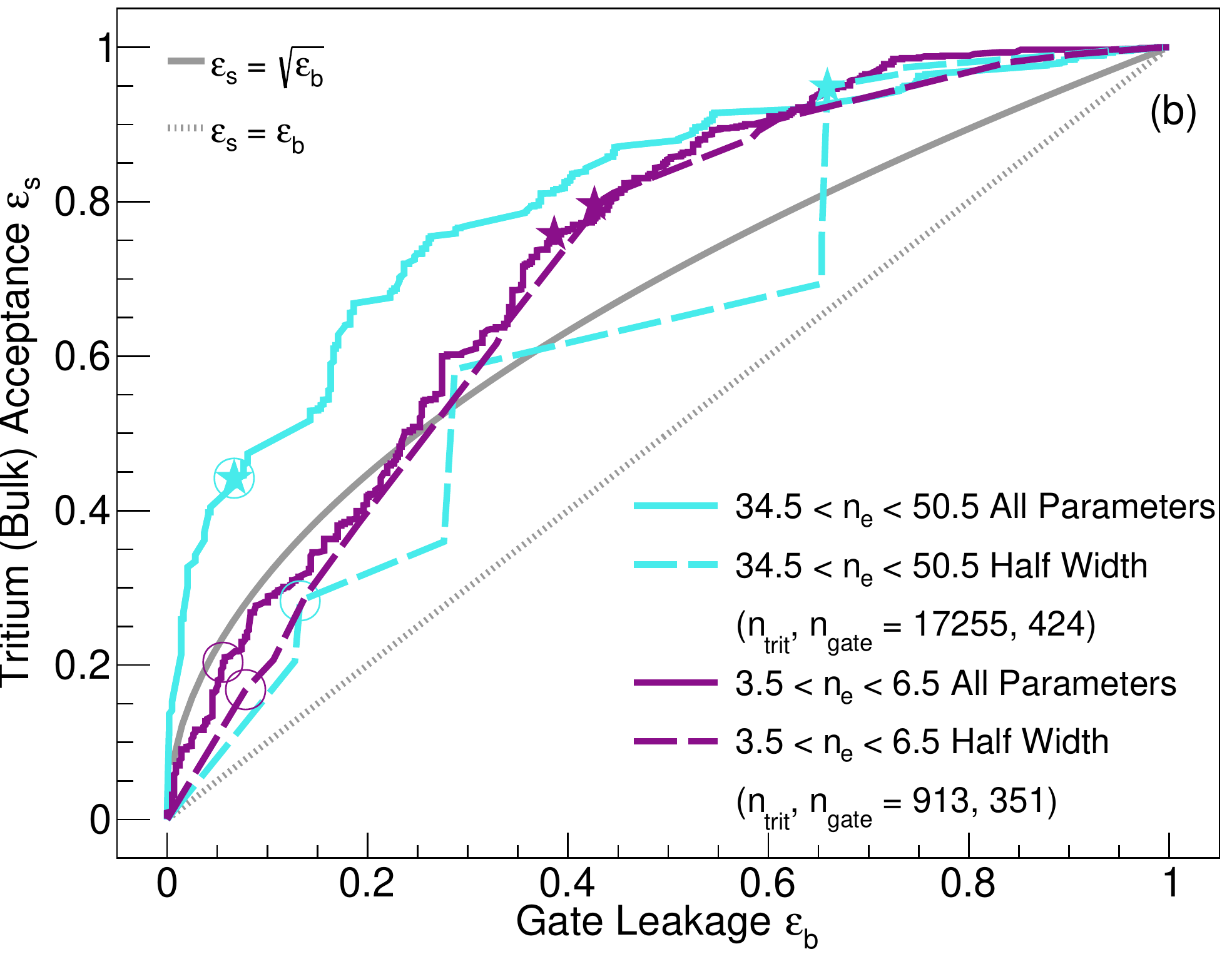}
  \caption{ROC curves of of training+testing data for a BDT using all shape quantifying parameters compared to a BDT using only half width. Points of maximum limit improvement are shown by stars and circles for background subtracted and Poisson cases, respectively.}
  \label{fig:roc_comparison}
\end{figure}

To characterize improvement of limits from using LUX's full set of shape quantifying parameters compared to just half width, a second BDT was run using half width as the only input parameter. Fig. \ref{fig:roc_comparison} compares the ROC curves of the ``total'' and ``half width'' BDTs for the smallest and largest $S2$ size bins using the combined training+testing dataset. For the cathode, both bins have similarly shaped total and half width ROC curves suggesting diffusion experienced by cathode $S2$s has the effect of washing away pulse shape information other than pulse width. For the gate, the ROC curves of the smallest bin show the same lack of improvement. However, in the largest bin, the total ROC curve sits far above the half width ROC curve indicating a large improvement from the addition of extra shape parameters.

The points of maximum improvement for background subtracted and Poisson limits are indicated as stars and circles in Fig.~\ref{fig:roc_comparison}. Poisson limits are maximized at equal or lower $\epsilon_s$ than background subtracted limits. Because they have a greater dependence on $\epsilon_b$ than background subtracted limits, Poisson limits favor lower values of $\epsilon_b$ despite the accompanying decrease in $\epsilon_s$.

\begin{figure}[ht]
  \includegraphics[width=\linewidth]{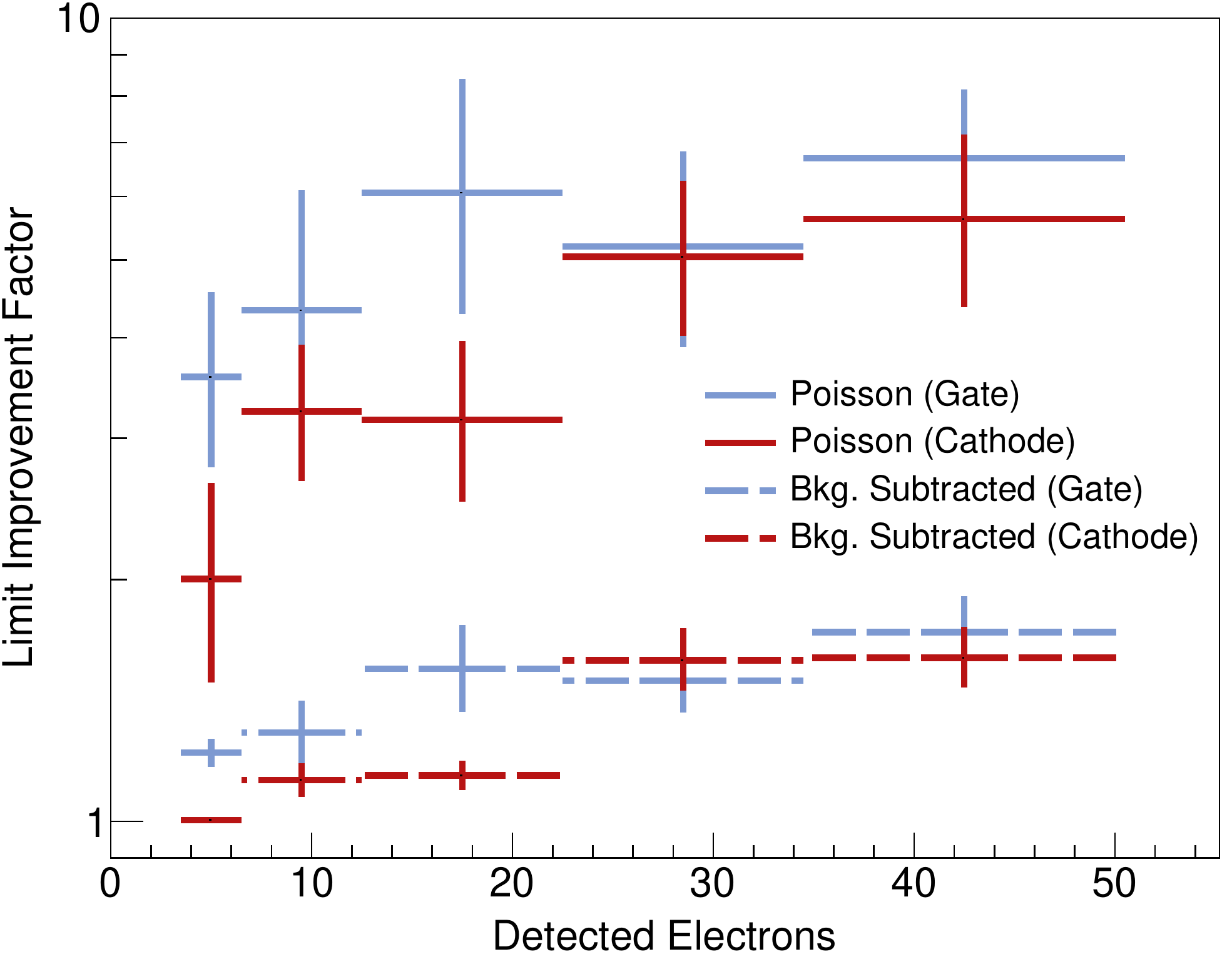}
  \caption{Predicted maximum improvement in exclusion limits for background subtracted ($\epsilon_s/\sqrt{\epsilon_b}$) and Poisson ($\epsilon_s/\epsilon_b$) scenarios, calculated separately for extremal cases of only gate backgrounds or only cathode backgrounds. Points of maximum improvement correspond to the stars and circles from Fig.~\ref{fig:roc_comparison} for the BDT using all parameters. Since we do not know the energy spectrum of the background, distinct values are given for each S2 size bin; the final improvement would be a weighted average of the values shown.}
  \label{fig:improvement}
\end{figure}

Figure~\ref{fig:improvement} uses the combined training+testing dataset to show the effectiveness of the BDT as it depends on $S2$ size. It shows the maximum expected limit improvement for each $S2$ size bin considering extremal cases of only gate backgrounds or only cathode backgrounds. Poisson limit improvement factors are in the range $2$--$7$ demonstrating our machine learning technique is extremely effective at improving limits based solely on comparing signal and observed event spectra. Poisson limit improvement is a factor of $2$--$3$ greater than that calculated for background subtracted limits, which can be explained by the Poisson limits' greater dependence on $\epsilon_b$. For all background scenarios, the limit improvement increases by a factor of $2$--$3$ from smallest to largest bin. An increase is expected due to the greater amount of pulse shape information encoded in larger pulses.

\section{Results and Sensitivity}\label{sec:results}
\begin{figure}[ht]
  \includegraphics[width=\linewidth]{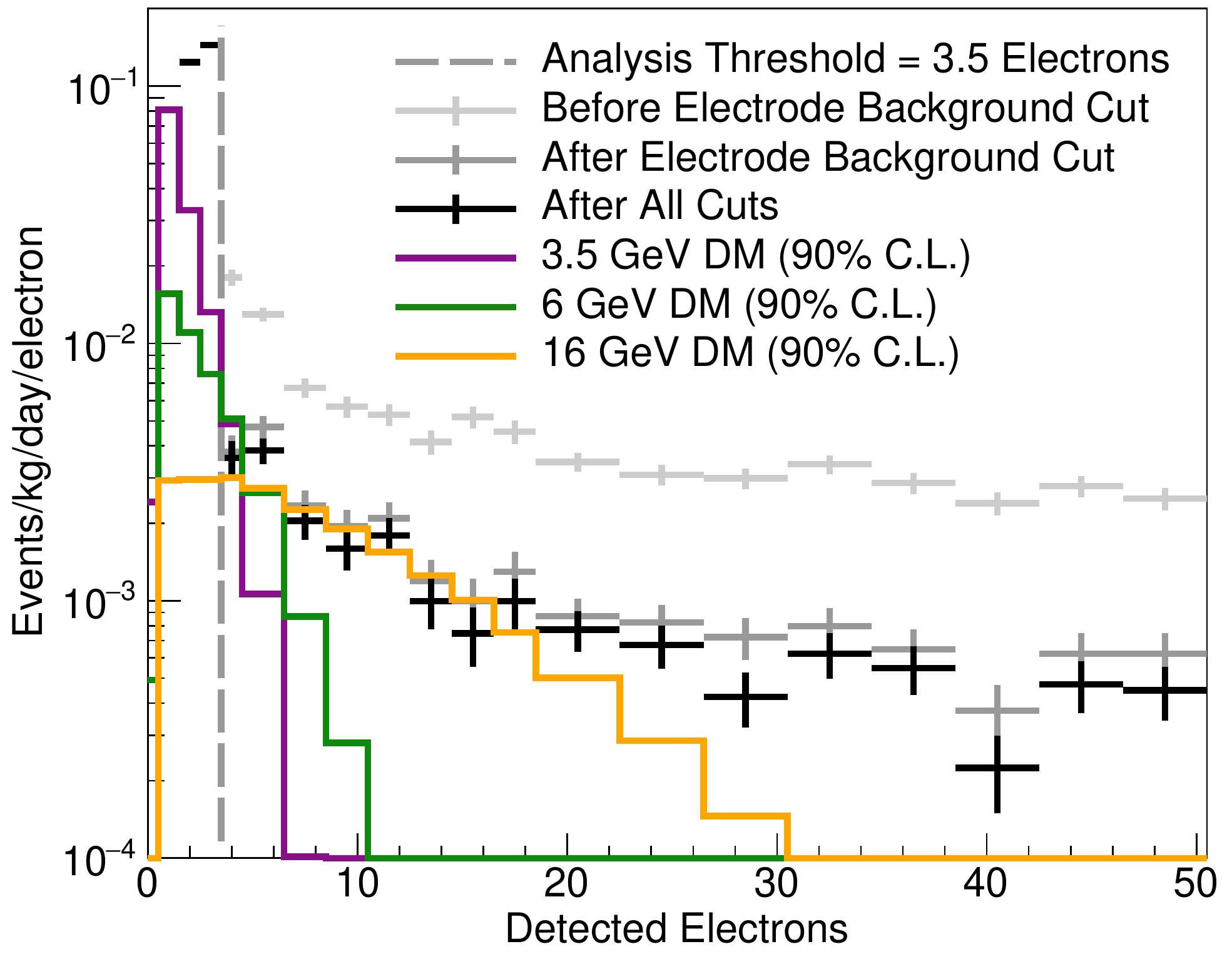}
  \caption{Dark matter (DM) search data from May through Sept 2013 before and after applying the BDT cut and hand-scanning to remove events that originated on the electrodes. The BDT cut reduces the observed event rate by a factor of $\sim4$ while retaining approximately $60$\% signal efficiency independent of $S2$ size. DM spectra at the $90$\% confidence interval cross-section limit are overlaid for comparison.}
  \label{fig:Spectrum}
\end{figure}

The final WS2013 data, after applying all cuts described in Sec.~\ref{sec:data-selection} and the BDT discriminator cut in Sec.~\ref{sec:electrode-backgrounds}~(A), are plotted in Fig.~\ref{fig:Spectrum}. The spectrum rises near the analysis threshold, though this is significantly mitigated by the BDT cut, which reduces the observed event rate by a factor of $\sim4$ while retaining approximatly $60$\% signal efficiency independent of $S2$ size. This outcome demonstrates a substantial and efficient removal of electrode backgrounds using only $S2$ pulse shape information, which is useful even for small $S2$s of only a few detected electrons. The residual rate of events just above the $3.5$ detected electron software threshold is approximately $7$~events/tonne/day/electron, after correcting for cut efficiencies.

\begin{figure}[ht]
  \includegraphics[width=\linewidth]{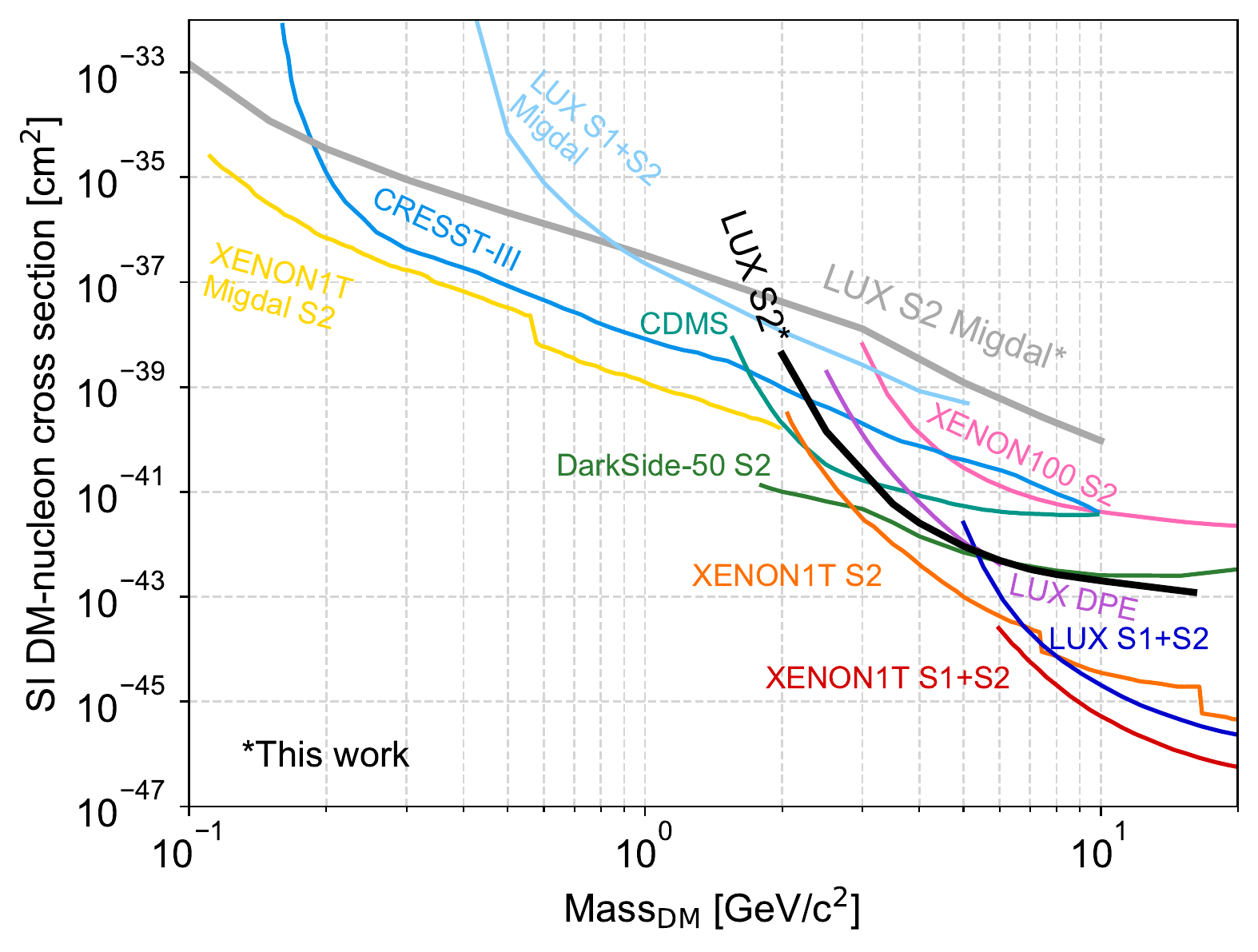}
  \caption{Upper limits on the spin-independent DM-nucleon cross section at $90$\% C.L. The result of the $S2$-only analysis with an NR signal model is shown in black, and the result of the $S2$-only analysis with a signal model based on the Migdal effect is shown in grey. Also shown are limits from DarkSide-50 \cite{DarkSide2018} ($S2$-only, binomial fluctuation assumption), CDMSlite \cite{CDMSLite2019}, CRESST-III \cite{CRESST-III2019}, XENON100 $S2$-only \cite{XENON100S2o2016}, XENON1T $S1+S2$ \cite{XENON1T2018}, XENON1T $S2$-only \cite{XENON1TS2o2019} (NEST 2.0.1 yields), XENON1T $S2$-only with Migdal effect \cite{XENON1TMigdal2019}, and past LUX searches using $S1+S2$ events \cite{LUX2016}, including S1s with single photons using double photoelectron emission (DPE) \cite{Akerib:2019zrt}, and the Migdal effect \cite{LUXMigdal2018}.}
  \label{fig:limits}
\end{figure}

Two DM signal hypotheses were tested against the data presented in Fig.~\ref{fig:limits}. First was the traditional spin-independent (SI) elastic scatter of DM on a xenon nucleus whereafter the nucleus recoils, ionizing and exciting neighboring xenon atoms in its path. Detection through this NR channel is limited to dark matter particles with $m_{\chi}\gtrsim2$~GeV which are able to transfer enough momentum to a heavy xenon nucleus to produce $S2$s above the analysis threshold. In a small fraction of DM scatters a second signal type is expected to be produced via the ``Migdal effect'' \cite{Ibe2018, Dolan2018}. It arises when the recoiling xenon nucleus induces a change in atomic energy levels, forcing the emission of a $\sim$keV electron that also ionizes and excites neighboring atoms. Although rare, this type of ER signal enables DM detectors to probe lower DM masses than those possible via the traditional NR channel. A small fraction of sub-GeV DM scatters are calculated to produce an above-threshold $S2$ via the Migdal effect, even when the NR signal is undetectable. While there has been no experimental confirmation of the Migdal effect to date, various experiments have applied the theory in DM searches \cite{EDELWEISS2019, CDEX1B2019, XMASS2019, XENON1TMigdal2019}. Details of the Migdal signal model applied in this work are published in \cite{LUXMigdal2018}, which outlines a search for DM via this channel using WS2013 data with both an $S1$ and $S2$.

\begin{figure}[ht]
  \includegraphics[width=\linewidth]{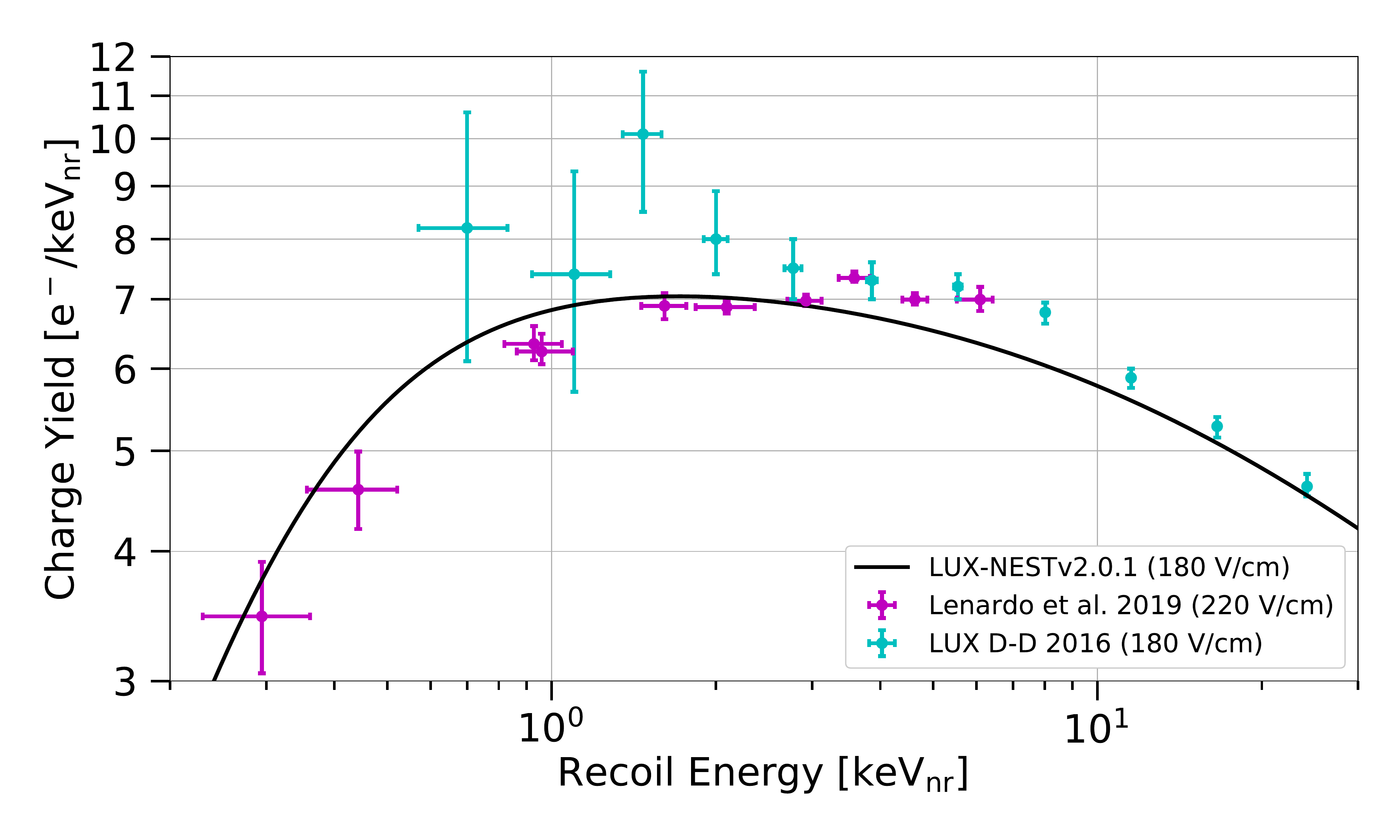}
  \includegraphics[width=\linewidth]{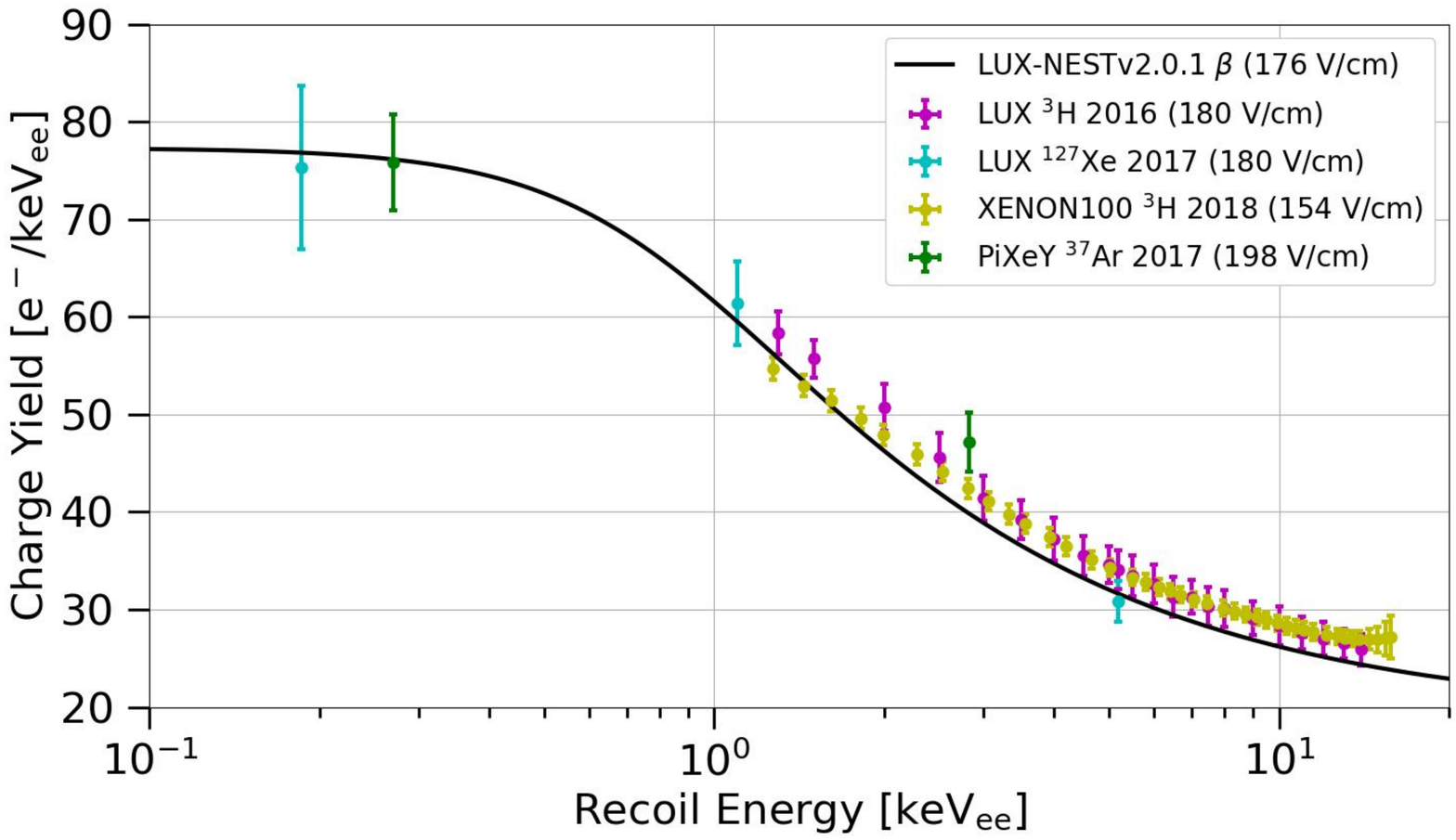}
  \caption{Sensitivity calculations employed the NEST model v2.0.1 NR and ER charge yield models (solid black curves), with a hard cutoff in yield below $0.3$~keV NR and $0.186$~keV ER.}
  \label{fig:Qy}
\end{figure}

The detectors responses to the traditional NR and Migdal signals were modeled with NEST v2.0.1---assuming a DM velocity distribution calculated from the Standard Halo Model as in \cite{PhysRevD.82.023530}. NEST relies on NR and ER charge yield models that are fit to empirical data, as shown in Fig.~\ref{fig:Qy}. Below the lowest experimental datapoints at $0.3$~keV$_{nr}$ and $0.186$~keV$_{ee}$, the charge yields are conservatively assumed to be zero. The overall signal efficiency is modeled according to Fig.~\ref{fig:Efficiency}, with an additional $\sim60$\% reduction from the BDT cut.

NR and Migdal upper limits presented in Fig.~\ref{fig:limits} were calculated using Yellin's $p_\mathrm{max}$ test statistic \cite{Yellin:2002xd}. The limits are an un-binned comparison between the set of WS2013 events passing all cuts and a signal model of one variable, in this case the $S2$ size. Unlike a simple Poisson analysis, this type of limit utilizes the difference in shape between the signal and observed spectra allowing for stronger exclusion limits without requiring a known background model.

A potential source of uncertainty in the signal model is introduced by the electrode cut. Because the gate and cathode are at the top and bottom of the detector, the cut has a greater tendency to remove signal-like events at short and long drift times compared to those at intermediate times. This can, in principle, affect the signal spectrum through the dependence of $S2$ size on drift time. This uncertainty was investigated by generating two versions of the signal spectra: one uniform in drift time and a second generated according to a Gaussian distribution narrowly-focused at the center of the detector. The limits generated by these two models differ by less than $4$\% at all DM masses, a negligible change. The curve in Fig.~\ref{fig:limits} is produced by the uniform distribution.

The low-threshold, $S1$-agnostic analysis described in this paper provides increased sensitivity to low-mass DM, compared to LUX's standard $S1+S2$ analyses of NR and Migdal signals \cite{Akerib:2016vxi,LUXMigdal2018}. This improvement is primarily due to inclusion of events with $S2$s smaller than $10$ detected electrons and no $S1$s. In this signal regime, one can see from Fig.~\ref{fig:Acceptance_Rejection} that gate events are rejected with greater efficiency than those from the cathode.

Through a combination of lower threshold, reduced inherent detector backgrounds, and careful removal of electrode events; LUX has been able to substantially improve on previous DM-nucleon scattering cross section limits from XENON100 \cite{XENON100S2o2016}, an experiment with a target mass similar to LUX. The present sensitivity is approximately comparable to that obtained by the DarkSide-50 experiment \cite{Agnes:2018ves}, though we note that the DarkSide-50 analysis relies on a broad spectrum energy calibration in the crucial regime below about $7$~keV. Other measurements in this regime suggest an approximately 30\% lower electron yield \cite{Tanaka_2020,Kimura_2020} that lessens with decreasing recoil energy. New, direct measurements are urgently needed to confirm those results. Finally, we note that the present results are not as stringent as those obtained by the larger XENON1T experiment \cite{XENON1TS2o2019}, whose exposure is approximatly a factor of $4$ greater than that of LUX.

Electrode events are likely to remain a challenge for LZ and other xenon experiments searching for low-mass dark matter, as well as coherent neutrino-nucleus scattering of solar $^8$B neutrinos. Because the machine learning technique introduced in this paper is highly dependent on $S2$ size, it is expected to be of equal or greater success in LZ due to low-energy signals containing greater numbers of electrons (primarily a result of a greater extraction efficiency, predicted to be $95$\% in LZ compared to $49$\% in LUX WS2013). The technique works in concert with treatments used to reduce electron emission rates from LZ grid wires. In particular, acid passivation was demonstrated by \cite{Tomas:2018pny} to bring about order-of-magnitude reductions of emission rate by improving the quality of the oxide layer on wire surfaces.

Finally, we suggest the success of the machine learning technique might be improved further by feeding entire $S2$ waveforms from each PMT into a convolutional neural network or similar algorithm intended for low level input, instead of just 12 shape-defining parameters. We note that such an approach would introduce additional computing burden, but it would likely lead to a significant improvement in discrimination power considering the substantial increase in potentially-useful information.

\section{Acknowledgments}
This work was partially supported by the
U.S. Department of Energy (DOE) under Awards
No. DE-AC02-05CH11231, DE-AC05-06OR23100,
DE-AC52-07NA27344, DE-FG01-91ER40618, DEFG02-08ER41549, DE-FG02-11ER41738, DEFG02-91ER40674, DE-FG02-91ER40688, DE-FG02-
95ER40917, DE-NA0000979, DE-SC0006605, DESC0010010, DE-SC0015535, and DE-SC0019066; the
U.S. National Science Foundation under Grants No.
PHY-0750671, PHY-0801536, PHY-1003660, PHY1004661, PHY-1102470, PHY-1312561, PHY-1347449,
PHY-1505868, and PHY-1636738; the Research Corporation Grant No. RA0350; the Center for Ultra-low
Background Experiments in the Dakotas (CUBED);
and the South Dakota School of Mines and Technology (SDSMT). Laborat\'{o}rio de Instrumenta\c{c}\~{a}o
e F\'{i}sica Experimental de Part\'{i}culas (LIP)-Coimbra
acknowledges funding from Funda\c{c}\~{a}o para a Ci\^{e}ncia
e a Tecnologia (FCT) through the Project-Grant
PTDC/FIS-NUC/1525/2014. Imperial College and
Brown University thank the UK Royal Society for
travel funds under the International Exchange Scheme
(IE120804). The UK groups acknowledge institutional
support from Imperial College London, University
College London and Edinburgh University, and from
the Science \& Technology Facilities Council for
Grants ST/K502042/1 (AB), ST/K502406/1 (SS), and
ST/M503538/1 (KY). The University of Edinburgh is a
charitable body, registered in Scotland, with Registration No. SC005336.
This research was conducted using computational
resources and services at the Center for Computation
and Visualization, Brown University, and also the Yale
Science Research Software Core.
We gratefully acknowledge the logistical and technical
support and the access to laboratory infrastructure provided to us by SURF and its personnel at Lead, South
Dakota. SURF was developed by the South Dakota
Science and Technology Authority, with an important
philanthropic donation from T. Denny Sanford. Its
operation is funded through Fermi National Accelerator
Laboratory by the Department of Energy, Office of High
Energy Physics.

\bibliography{main}
\end{document}